\documentclass[10pt,aps,prd,amsmath,amssymb,superscriptaddress,twocolumn,nofootinbib,showpacs,preprintnumbers,amsfonts,notitlepage]{revtex4-1}

\usepackage{XYZ}
\usepackage{orcidlink}
\usepackage{affiliations}

\allowdisplaybreaks

\begin{document}

\preprint{JLAB-THY-24-4009}

\title{$XY\!Z$ spectroscopy at electron-hadron facilities III: \\
       Semi-inclusive processes with vector exchange}
\author{D.~\surname{Winney}\orcidlink{0000-0002-8076-243X}}
\email{winney@hiskp.uni-bonn.de}
\affiliation{\HISKP}
\author{A.~\surname{Pilloni}\orcidlink{0000-0003-4257-0928}}
\email{alessandro.pilloni@unime.it}
\affiliation{\messina}
\affiliation{\catania}
\author{R.~J.~\surname{Perry}\orcidlink{0000-0002-2954-5050}}
\affiliation{\ub}
\author{\L.~\surname{Bibrzycki}\orcidlink{0000-0002-6117-4894}}
\affiliation{\AGH}
\author{C.~\surname{Fern\'andez-Ram\'irez}\orcidlink{0000-0001-8979-5660}}
\affiliation{\uned}
\author{N.~\surname{Hammoud}\orcidlink{0000-0002-8395-0647}}
\affiliation{\ifj}
 \author{V.~\surname{Mathieu}\orcidlink{0000-0003-4955-3311}}
 \affiliation{\ub}
\author{G.~\surname{Monta\~na}\orcidlink{0000-0001-8093-6682}}
\affiliation{\jlab}
\author{A.~\surname{Rodas}\orcidlink{0000-0003-2702-5286}}
\affiliation{\jlab}
\affiliation{\odu}
\author{V.~\surname{Shastry}\orcidlink{0000-0003-1296-8468}}
\affiliation{\ceem}
\affiliation{\indiana}
\author{W.~A.~\surname{Smith}\orcidlink{0009-0001-3244-6889}}
\affiliation{\gwu}
\affiliation{\ucb}
\affiliation{\lbnl}
\author{A.~P.~\surname{Szczepaniak}\orcidlink{0000-0002-4156-5492}}
\affiliation{\jlab}
\affiliation{\ceem}
\affiliation{\indiana}

\collaboration{Joint Physics Analysis Center}

\begin{abstract}
Inclusive production processes will be important for the first observations of $XYZ$ states at new generation electron-hadron colliders, as they generally benefit from larger cross sections than their exclusive counterparts.  We make predictions of semi-inclusive photoproduction of the $\chi_{c1}(1P)$ and $X(3872)$, whose peripheral production is assumed to be dominated by vector exchanges. 
We validate the applicability of Vector Meson Dominance in the axial-vector charmonium sector and calculate production rates at center-of-mass energies relevant for future experimental facilities.
We find the semi-inclusive cross sections near threshold to be enhanced by a factor of $\sim 2\text{--}3$ compared to the exclusive reaction and well suited for a first observation in photoproduction.
\end{abstract}

\date{\today}
\maketitle

\section{Introduction}
\label{sec:Intro}

The discovery of \XYZ signals in the quarkonium sector have stimulated much theoretical and experimental research.
In order to investigate their structure, it is first necessary to confirm that these phenomena are genuine hadronic resonances. The majority of them have been observed only in specific reactions and searches in new production modes, are required to confirm their status in the hadron spectrum, and to provide us with complementary information on their internal dynamics~\cite{Guo:2019twa,JPAC:2021rxu,Esposito:2016noz,Guo:2017jvc,Brambilla:2019esw}. In particular, it has been argued that the coupling of the $X(3872)$ to photons could distinguish it from an ordinary charmonium with the same quantum numbers, like the $\chi_{c1}(1P)$~\cite{Babiarz:2022xxm,Babiarz:2023ebe}, and thus photon-induced production is a promising avenue for $\XYZ$ spectroscopy.

In the first of a series of papers~\cite{Albaladejo:2020tzt}, we studied the exclusive photoproduction of several \XYZ states. 
Exclusive reactions have the advantage of having well-constrained kinematics, and generally cleaner experimental signals. In a subsequent publication~\cite{Winney:2022tky}, we extended the predictions to semi-inclusive production. While such processes are generally expected to suffer from larger backgrounds, they benefit from higher cross sections. This latter analysis was restricted to states generated via charged pion exchange, that is, to the production of $Z_c^\pm$ and $Z_b^{(\prime)\pm}$. The predicted cross sections indicate that these states can be amply produced in next-generation electron-hadron facilities~\cite{AbdulKhalek:2021gbh,Anderle:2021wcy,Accardi:2023chb}. 

Searches for the $X(3782)$ have been proposed for a variety of different inclusive hadroproduction reactions which may give insight into its 
composition~\cite{Bignamini:2009sk,*Bignamini:2009fn,*Artoisenet:2010uu,*Artoisenet:2009wk,*Albaladejo:2017blx,*Esposito:2017qef,*Zhang:2020dwn,*Esposito:2020ywk,*Braaten:2020iqw}. For semi-inclusive electroproduction, Ref.~\cite{Yang:2021jof} predicts production rates within the molecular model of the $X(3872)$ emerging from short-distance production of $D\bar{D}^*$ pairs which then rescatter. Because these latter predictions are based on the perturbative parton picture encoded in PYTHIA~\cite{Sjostrand:2006za}, they may not be reliable for \mbox{(quasi-)real} photoproduction, where other production mechanisms may also contribute. The goal of the present paper is to continue our exploration of semi-inclusive production to focus on vector exchanges, and thus the production of the $X(3872)$.

Vector exchanges are relevant even for the conventional $\chi_{c1}(1P)$ which has been studied in Refs.~\cite{Benic:2024pqe,Jia:2022oyl}. In particular, photon exchange serves as a background to the possible odderon (\textit{aka} 3-gluon~\cite{Brodsky:2000zc}) exchange in high-energy photoproduction. Odderon exchanges will not be considered in this work, but could enhance the axial meson production over the rates predicted here.

Predictions for the photoproduction of \XYZ states are complicated by the lack of knowledge of their radiative decay into light mesons. In order to estimate the required photocouplings, a common approach is to employ vector meson dominance (VMD)~\cite{Sakurai:1960ju}, where typically the photon is replaced with a sum of ground-state vector mesons, each weighted by a coupling constant determined from their electronic widths. Models based on VMD have been applied with some success in the phenomenology of light mesons (see e.g. \cite{Meissner:1987ge} and references therein) and have also been employed in the study of heavy quarkonia, in cases where suitable experimental observables have not yet been measured.
The applicability of these models, however, has been criticized recently~\cite{Du:2020bqj,Cao:2024nxm,Xu:2021mju}, especially when heavy mesons are involved. Further, a recent analysis~\cite{JointPhysicsAnalysisCenter:2023qgg} suggests that VMD-based production models fail to describe $J/\psi$ photoproduction data near threshold~\cite{GlueX:2023pev,Duran:2022xag} and raises questions about the continued application of VMD to heavy quarkonia.

As noted in Ref.~\cite{JPAC:2021rxu}, however, there are cases even in the charmonium sector where measurements are roughly consistent with VMD expectations. This observation suggests that the applicability of VMD should be assessed on a case-by-case basis. We thus reexamine exclusive photoproduction of $\chi_{c1}(1P)$ and $X(3872)$ and show that the application of VMD leads to order-of-magnitude predictions consistent with measured decay widths and other established phenomenology.

Encouraged by this agreement in the exclusive case, we generalize the vector exchange mechanism to the semi-inclusive extension of the vector exchange mechanism final state. This closely mirrors the generalization of the pion exchange in Ref.~\cite{Winney:2022tky} but with additional complications introduced by exchanges of particles with spin. As in the scalar exchange case, the semi-inclusive generalization of a $t$-channel vector exchange only affects the target fragmentation and is thus agnostic to the nature of the particle $\mQ$. The total inclusive production rates for the $\chi_{c1}(1P)$ and $X(3872)$ can thus be predicted and are found to be a factor of $2\text{--}3$ enhancement over exclusive production alone in the near-threshold region. 
Because of this, semi-inclusive searches are very promising for a first observation at future facilities. 

The paper is organized as follows: In \cref{sec:ExVMD}, we revisit the exclusive production of $\chi_{c1}(1P)$ and $X(3872)$ via photon and vector meson exchanges.
In \cref{sec:SemiInclusive} we extend the analysis 
  to semi-inclusive production using inclusive proton structure functions. These functions are considered in two regions of interest separately: in the near-threshold/resonance region with small missing mass, and at larger center-of-mass (CM) energies, where missing mass can be large. In \cref{sec:numerical}, we discuss the numerical results for the cross sections relevant to future experimental facilities. Finally, concluding remarks and a summary are given in \cref{sec:Conclusions}.
\begin{figure}[t]
    \centering
    \includegraphics[width=.5\columnwidth]{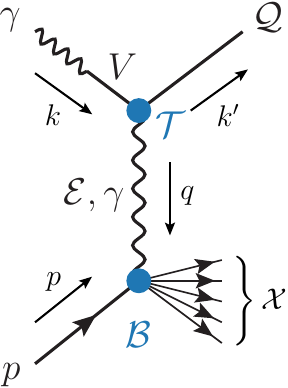}
    \caption{\label{fig:momlabels}    
     Semi-inclusive photoproduction of $J^{PC} = 1^{++}$ $\mQ$ quarkoniumlike state
via vector meson (\mE) or photon exchange~($\gamma$) in the $t$-channel. The bottom vertex $\mB$ is
generalized to consider the production of arbitrary final state
$\mX$. Compared to the notation of~\cite{Albaladejo:2020tzt,Winney:2022tky}, we switched the momentum of the external and exchanged bosons to be $k^{\prime}$ and $q$ respectively to have $t=q^2$. The exclusive case is recovered when $\mX$ is a single proton with momentum $p' = p + q$.}
\end{figure}

\section{Revisiting VMD in exclusive $X$ production}
\label{sec:ExVMD}

Thanks to the variety of measured decay widths, it is possible to test the applicability of VMD in the case of $\chi_{c1}(1P)$ and $X(3872)$. The production of $X(3872)$ has been measured in two-photon collisions as well as hadronic and radiative decays into vector mesons. For the $\chi_{c1}(1P)$, all its radiative decay widths have been observed. While no two-photon coupling has been measured for the $\chi_{c1}(1P)$, there exist several microscopic calculations that one can use for comparison. Since VMD relates all these processes, one can test to what degree data are actually consistent with VMD expectations. 
In order to concretely examine possible relations between photon and meson exchange quantities, we revisit the exclusive production and motivate a form of the amplitude which can be used for both massive and massless exchange particles. 
\subsection{Photon Exchange}
We first consider the exclusive production of an axial-vector meson $\mQ$ via a $t$-channel photon exchange, as represented in \cref{fig:momlabels} when $\mX$ is a single nucleon. The amplitude can be written as:
    \begin{equation}
        \label{eq:photon_exchange}
        \mel{\lambda_\gamma,\lambda_N}{T_\gamma}{ \lambda_\mQ \, \lambda_{N^\prime} } = \mathcal{T}_{\lambda_\gamma\lambda_\mQ}^{\mu} \, g_{\mQ\gamma\gamma} \left[\frac{-g_{\mu\nu}}{t}\right] e \, \mathcal{B}^{\nu}_{\lambda_N\lambda_{N^\prime}} ~,
    \end{equation}
in terms of factorized ``top" and ``bottom" tensors incorporating the $\mQ\gamma\gamma$ and $\gamma N N$ interactions respectively, as well as the photon propagator. For the top vertex, we choose a gauge-invariant interaction with a minimal number of derivatives:
    \begin{equation}
        \label{eq:NewLagrange}
        \mathcal{L}_{\mathcal{Q}\gamma\gamma} = \frac{1}{2} \, \frac{g_{\mathcal{Q}\gamma\gamma}}{m^2_\mathcal{Q}} \, \epsilon_{\alpha\beta\mu\nu} \, F^{\alpha\beta} \, \partial_\sigma \, F^{\sigma\mu} \, \mathcal{Q}^{\nu} ~,
    \end{equation}
where $F^{\mu\nu}$ is the photon field strength tensor and $\mathcal{Q}^\nu$ is the axial vector meson field. Note that, if both photons are on-shell, the interaction vanishes since $\partial_\mu F^{\mu\nu}=0$ as expected by the Landau-Yang theorem~\cite{Landau:1948kw,Yang:1950rg}. The Lagrangian implies:
    \begin{align}
        \label{eq:mT}
        \mathcal{T}_{\lambda_\gamma\lambda_\mathcal{Q}}^\mu &= \frac{1}{m_\mathcal{Q}^2} \, f_{\mathcal{Q}\gamma\gamma}(t) \, \epsilon_{\alpha\beta\sigma\nu}
        \\
        & \qquad \times k^\alpha \, \varepsilon^\beta(k, \lambda_\gamma) \, \left[ q^2 \, g^{\sigma\mu} - q^\sigma \, q^\mu \right] \varepsilon^{*\nu}(k^\prime, \lambda_\mQ) ~,
        \nonumber
    \end{align}
where we factor out the couplings $g_{\mQ\gamma\gamma}$ in \cref{eq:photon_exchange} and include an additional form factor 
    \begin{equation}
        \label{eq:Fqgg}
        f_{\mQ\gamma\gamma}(t) = \frac{1}{1-t/m_\mQ^2} ~,
    \end{equation}
which parameterizes the finite size of the top vertex. The scale is chosen as the quarkonium mass $m_\mQ$, based on the model of~\cite{Schuler:1997yw} for two-photon transition form factors of $C$-even charmonia, and will compensate unphysical polynomial growth of \cref{eq:NewLagrange} at large $t$.

\begin{figure}[t]
        \centering
        \includegraphics[width=\columnwidth]{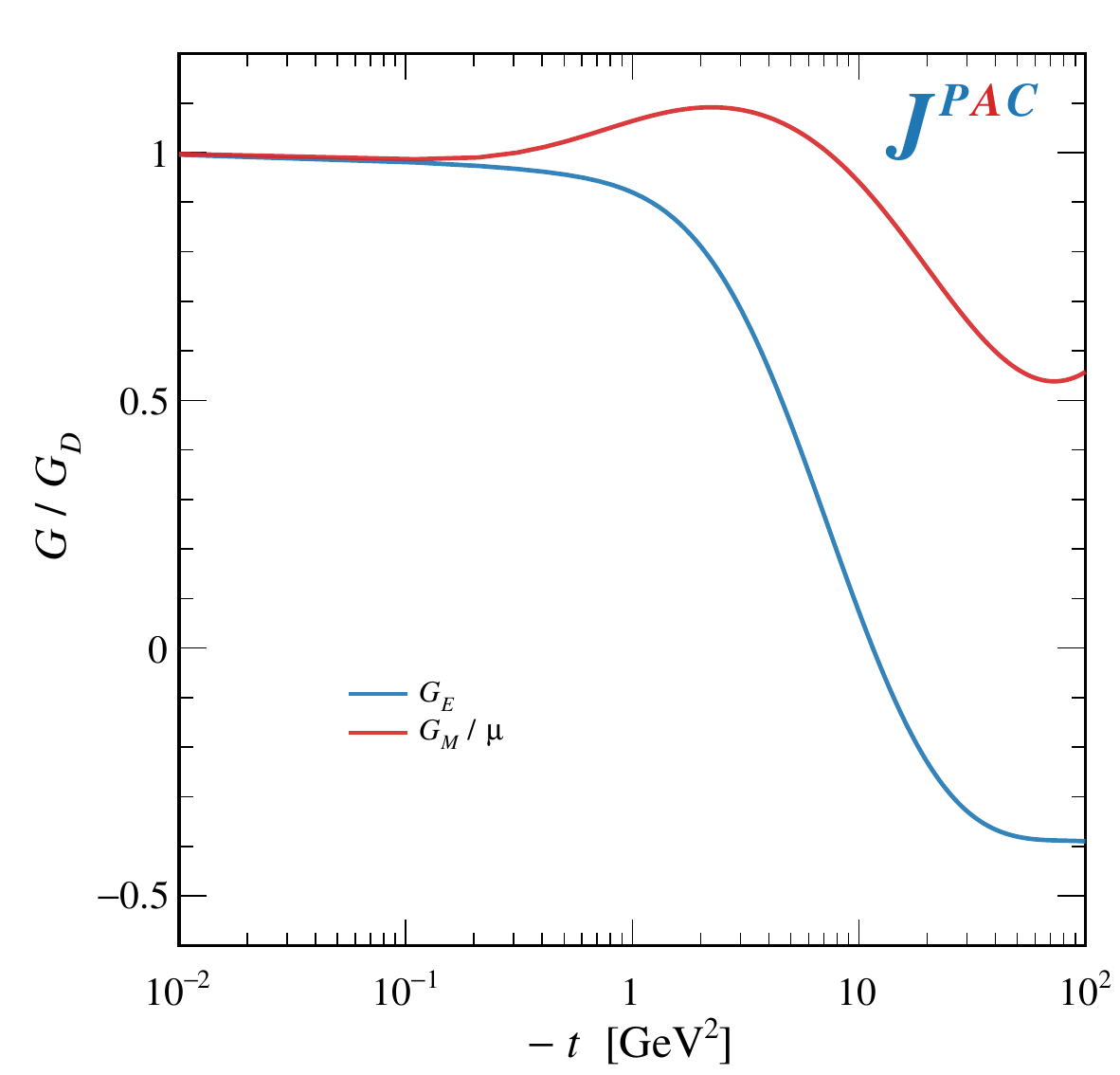}
        \caption{Sachs electric and magnetic form factors for the proton using the parameterization of Ref.~\cite{Ye:2017gyb}. For the magnetic form factor, we divide by the magnetic moment of the proton, $\mu = 2.79$.}
        \label{fig:sachsFFs}
    \end{figure}

The bottom vertex describes the $\gamma N N$ interaction 
\begin{subequations}
    \label{eq:mB}
    \begin{align}
        \mathcal{B}_{\lambda_N,\lambda_{N^\prime}}^\mu =  \bar{u}(p^\prime, \lambda_{N^\prime}) \, \Gamma^\mu(q)  \, u(p, \lambda_N) ~,
    \end{align}
and is parameterized in terms of the Dirac and Pauli form factors~\footnote{$F_1(0) = 1$ meanwhile $F_2(0) = 1.79$ is normalized to the anomalous magnetic moment of the proton.}, $F_{1,2}$, which are functions of the exchanged photon virtuality $t = q^2$,
    \begin{equation}
        \label{eq:Gammaq_photon}
        \Gamma^\mu(q) = \bigg[F_1 \, \gamma^\mu + F_2 \, \frac{i\sigma^{\mu\nu}q_\nu}{2 \, m_N} \bigg] ~.
    \end{equation}
\end{subequations}
 We further write these form factors in terms of the Sachs electric and magnetic form factors $G_{E,M}$~\cite{Punjabi:2015bba}:
    \begin{equation}    
        \label{eq:Pauli&DiracFs}
        F_1 = \left(\frac{G_E + \tau \, G_M}{1 + \tau}\right)
        \quad 
        F_2 = \left(\frac{G_M - G_E}{1 + \tau}\right) ~,
    \end{equation}
where $\tau \equiv -t / 4\,m_N^2$. In numerical calculations we use the parameterization of Ref.~\cite{Ye:2017gyb} for $G_{E,M}$ which
in  \cref{fig:sachsFFs}  are shown normalized to the dipole form factor
    \begin{equation}
        \label{eq:G_D}
        G_D(t) = \left(1 + \frac{-t}{0.71\gevsq}\right)^{-2} . 
    \end{equation}
 These electromagnetic form factors are known to high precision for both the proton and neutron. In principle, this knowledge allows us to predict production cross sections for targets of both isospins. Photoproduction off a neutron target is experimentally more difficult and produces smaller rates, therefore we will only show results for the proton target but code to produce plots for neutron targets are available online~\cite{zenodo}.
   \begin{table}[b]
        \caption{Couplings of the bottom vertex for the vector exchanges from~\cite{Albaladejo:2020tzt} and references therein.
        }
        \begin{ruledtabular}
        \begin{tabular}{c c cc}
            $\mE$ &             $g_{\mE NN}$ & $g^\prime_{\mE NN}$ & $\Lambda_\mE$ [\gevp]
            \\
            \hline
            $\rho$ & $2.4$ & $14.6$ & 1.4\\
            $\omega$ &  $16$ & $0$ &  $1.2$
        \end{tabular}
        \end{ruledtabular}
        \label{tab:X_nucleon_couplings}
    \end{table}
\subsection{Meson Exchange}

We now consider the analogous process to \cref{eq:photon_exchange} with the exchange photon replaced by a massive vector meson. The meson exchange process was studied in Ref.~\cite{Albaladejo:2020tzt}, with the top vertex parameterized using an effective Lagrangian of the form
    \begin{equation}
        \label{eq:OldLagrange}
        \tilde{\mathcal{L}}_{\mQ\gamma\mE} = \frac{1}{2} \, \tilde{g}_{\mQ\gamma\mE} \,
        \epsilon_{\alpha\beta\mu\nu} \, F^{\alpha\beta} \, \mE^{\mu} \, \mQ^{\nu} ~,
    \end{equation}
where $\mE^\nu$ denotes the vector meson field. We instead use the same Lagrangian as in \cref{eq:NewLagrange}, simply replacing the photon by the vector field strength tensor,\footnote{This effectively imposes an additional $U(1)$ symmetry on the interaction with respect to the massive exchange particle, which is analogous to the St\"{u}ckelberg construction for massive electrodynamics~\cite{Ruegg:2003ps,Meissner:1987ge}.} yielding the same top tensor as in \cref{eq:mT}. By doing so, we also include the charmonium-to-$\gamma\gamma$ form factor of \cref{eq:Fqgg} which was not considered in our previous study. Its presence will not affect numerical results significantly, as momentum transfers above $\sim1\gevsq$ will be cut off by meson-nucleon form factors. 

Note that $\tilde{\mathcal{L}}_{\mQ\gamma\mE}$ and $\mathcal{L}_{\mQ\gamma\mE}$ yield the same on-shell amplitude if the couplings are related by:
    \begin{equation}
        \label{eq:coupling_redef}
        g_{\mQ\gamma\mE} = \left(\frac{m_\mQ}{m_\mE}\right)^2 \, \tilde{g}_{\mQ\gamma\mE} ~.
    \end{equation}
The advantage of using \cref{eq:NewLagrange,eq:coupling_redef} is that we may use the Feynman gauge propagator also for the massive exchange:
    \begin{equation}
        \label{eq:propagator}
        \mP^{\mu\nu}_\mE = \frac{-g^{\mu\nu}}{t - m_\mE^2} ~,
    \end{equation}
and therefore the same functional form for both photon and meson exchange amplitudes. 
The vector meson exchange amplitude, may thus be written as:
    \begin{align}
        \label{eq:explicit_meson}
        \mel{\lambda_\gamma,\lambda_N}{\tilde{T}_\mE}{ \lambda_\mQ \, \lambda_{N^\prime} }& 
        \nonumber \\
        = \mathcal{T}_{\lambda_\gamma\lambda_\mQ}^{\mu} \, g_{\mQ\gamma\mE}&\left[\frac{-g_{\mu\nu}}{t- m_\mE^2}\right] \, g_{\mE NN} \, \tilde{\mathcal{B}}^{\nu}_{\lambda_N\lambda_{N^\prime}}
         ~.
    \end{align}
The bottom vertex here is defined in terms of meson-nucleon couplings and form factors:
    \begin{equation}
        \label{eq:Gammaq_meson}
        \tilde{\Gamma}_\mE^{\mu}(q) = e^{t^\prime/\Lambda^2_\mE} \, \bigg[ \gamma^\mu + \frac{g^\prime_{\mE NN}}{g_{\mE NN}} \, \frac{i\sigma^{\mu\nu}q_\nu}{2 \, m_N} \bigg] ~,
    \end{equation}
with $t' = t - t_\text{min} = t - t(s, \cos\theta_s = 1)$. The values of the couplings $g^{(\prime)}_{\mE NN}$ and the cutoff parameters $\Lambda_\mE$ are listed  in \cref{tab:X_nucleon_couplings} and taken from Ref.~\cite{Albaladejo:2020tzt} and references therein.

\subsection{Applying VMD}
VMD predicts relations between hadronic and electromagnetic couplings~\cite{Sakurai:1960ju}, 
\begin{subequations}
    \label{eq:VMD_couplings}
    \begin{align}
        \label{eq:VMD_couplings_photon}
        g_{\mQ \gamma\gamma} &= \sum_i \frac{g_{\mQ i\gamma}}{\gamma_i}~, \\
      \label{eq:VMD_couplings_radiative}
        g_{\mQ \gamma\mE} &= \sum_{V} \frac{g_{\mQ V \mE}}{\gamma_V}~,\\
        \label{eq:VMD_couplings_nucleon}
         g_{\gamma NN} &= \sum_{\mE} \frac{g_{\mE NN}}{\gamma_\mE}  ~,
    \end{align}
\end{subequations}
where the sums generally run over all vector mesons, i.e. $\rho, \omega,\phi, \text{ and } \psi$. In the context of the reaction considered here, we use $\mE$ and $V$ to differentiate the vector mesons which are exchanged in the $t$-channel and coupled to the photon beam respectively. In \cref{eq:VMD_couplings_photon}, the sum may run over either particle and we use a generic label $i$.
The constants $\gamma_{i}$ are tabulated in \cref{tab:VMD_pars} and are determined by the decay constant $f_i$ of the vector meson:
    \begin{equation}
        \label{eq:eta}
        \frac{1}{\gamma_i} = \frac{e \, f_i}{m_i} ~.
    \end{equation}
    \begin{table}[b]
        \centering
        \caption{VMD parameters for different vector mesons. Decay constants are taken from~\cite{Mathieu:2018xyc,Albaladejo:2020tzt} and references therein.}
        \label{tab:VMD_pars}`
        \begin{ruledtabular}
        \begin{tabular}{ccccc}
            $i$ & $m_i$ [\mevp] &  $f_i$ [\mevp] & $\gamma_{i}$ \\
            \hline
            $\rho$ & 775.26 &  156.38 & 16.37 \\
            $\omega$ & 782.65 & 45.87 & 56.34 \\
            $\phi$ & 1019.46 & 75.87 & 44.37 \\
            $J/\psi$ & 3096.90 & 277.40 & 36.85
        \end{tabular}
    \end{ruledtabular}
    \end{table}

The radiative decays of the meson $\mQ$ are related to the two-photon interaction in \cref{eq:VMD_couplings_photon} through VMD. This prediction thus provides a testable consistency relation when the couplings on both sides of the equality are known.

The radiative couplings needed for the top vertex are related to the purely hadronic couplings in  \cref{eq:VMD_couplings_radiative}. While the sum runs over all possible vector mesons, the produced state $\mQ$ contains a heavy $c\bar c$ pair, and its coupling to purely light final states is OZI suppressed and expected to be small. 
We may thus restrict the sum to only $V = J/\psi$ in \cref{eq:VMD_couplings_radiative} to write:
    \begin{equation}    
        \label{eq:VMD_top}
        g_{\mQ\gamma\mE} = \frac{g_{\mQ\psi\mE}}{\gamma_\psi} ~.
    \end{equation}

The hadronic couplings for the bottom vertex are analogously related in \cref{eq:VMD_couplings_nucleon} except here the opposite happens: since the nucleon contains only light quark constituents, its couplings to $\phi$ and $\psi$ are OZI-suppressed~\cite{Bijker:2004yu,Bijker:2005pe}. Furthermore, such exchanges would be additionally suppressed by their heavier masses in the exchange propagator and can be neglected. 
The remaining couplings to $\mE =\rho, \omega$ can be determined using isospin symmetry, which allows the sum in  \cref{eq:VMD_couplings_nucleon} to be inverted to obtain the hadronic couplings from the electromagnetic ones. Writing
\begin{subequations}
    \begin{align}
        g_{\gamma pp} &= \frac{g_{\omega NN}}{\gamma_\omega} + \frac{g_{\rho NN}}{\gamma_\rho} = e~, \\
         g_{\gamma nn} &= \frac{g_{\omega NN}}{\gamma_\omega} - \frac{g_{\rho NN}}{\gamma_\rho} = 0~,
    \end{align}
\end{subequations}
yields 
    \begin{equation}    
        \label{eq:VMD_bottom}
        g_{\mE NN} = \frac{\gamma_\mE}{2} \, e ~.
    \end{equation}
Motivated by this relation, we propose an alternative model to \cref{eq:explicit_meson} for the vector meson exchange amplitude by using VMD to rescale the electromagnetic bottom tensor in \cref{eq:mB}:
    \begin{align}
        \label{eq:vector_exchange}
        \mel{\lambda_\gamma,\lambda_N}{T_\mE}{ \lambda_\mQ \, \lambda_{N^\prime} } =&
        \\
        \quad\, \mathcal{T}_{\lambda_\gamma\lambda_\mQ}^{\mu} \, g_{\mQ\gamma\mE} \, \bigg[&\frac{-g_{\mu\nu}}{t- m_\mE^2}\bigg] \, \beta_\mE(t^\prime) \, \frac{\gamma_\mE }{2} \, e \, \mathcal{B}^{\nu}_{\lambda_N\lambda_{N^\prime}}
         \nonumber ~.
    \end{align}
Here the top coupling may be calculated directly from a radiative decay width or from the additional use of VMD at the top vertex with \cref{eq:VMD_top}.

For compatibility with the predictions previously made using \cref{eq:Gammaq_meson} and based on models of meson-nucleon interactions, we restore the exponential $t$-behavior with an additional form factor:
    \begin{equation}
        \label{eq:beta}
        \beta_\mE(t^\prime) = e^{t^\prime/\Lambda_\mE^2} / \, G_D(t^\prime) ~.
    \end{equation}
The dipole factor is divided out so as to not double count the suppression at intermediate values. \Cref{eq:beta} is parameterized to be equal to one at forward angles and with the same $\Lambda_\mE$ as in \cref{tab:X_nucleon_couplings}, such that the scale of the amplitude is still primarily dictated by the VMD couplings, i.e. through \cref{eq:VMD_bottom}.

\begin{figure}[t]
        \centering
        \includegraphics[width=\columnwidth]{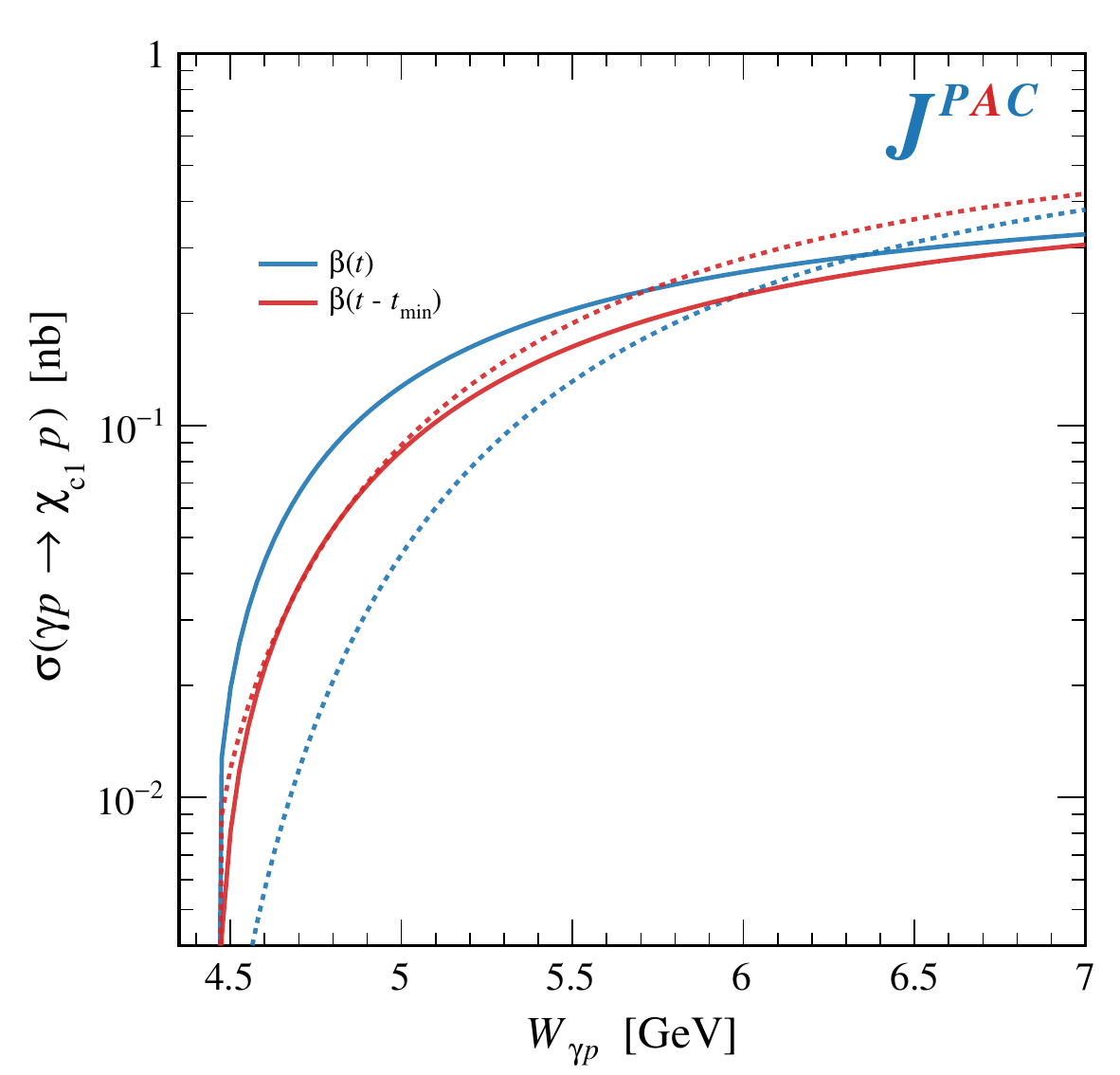}
        \caption{Comparison of the $\chi_{c1}(1P)$ production for different choices of argument for the exponential form factor in \cref{eq:beta}. Solid curves are calculated with VMD as in \cref{eq:vector_exchange}, while dashed curves are given by the original vector meson exchange as in \cref{eq:explicit_meson}.}
        \label{fig:FF_compare}
    \end{figure}

We chose the argument of the exponential form factor in \cref{eq:Gammaq_meson} to be $t^\prime$ instead of $t$ to avoid an unjustified suppression in the region close to threshold where $t_\text{min}$ is large. This, however, was an \textit{ad hoc} solution as it introduces a spurious $s$ dependence in the form factor which should in principle be only a function of the momentum transfer. To gauge the systematics of this choice, we plot the $\chi_{c1}(1P)$ production cross section for different choices of $t$ or $t'$ in the form factors in \cref{fig:FF_compare}. 
In the following, we will use $\beta(t^\prime)$ as in \cref{eq:beta} to be consistent with previous predictions.  

    \begin{table*}
    \caption{\label{tab:raw_couplings} Summary of ``top" vertex couplings (i.e. two-photon, radiative, and hadronic) which can be calculated directly from measured decay widths. All couplings are calculated with the interaction \cref{eq:mT} with $V$ and $\mE$ referring to the ``beam" and exchange particles respectively. The hadronic couplings are taken from \cite{Albaladejo:2020tzt} and redefined by \cref{eq:coupling_redef} as described in the text.}
    \begin{ruledtabular}
        \begin{tabular}{ccccccc}
            $\mQ$ & $V$  & $\mE$ & $g_{\mQ\gamma\gamma}^\text{(meas.)}$ & $g_{\mQ\gamma\mE}^\text{(meas.)}$ & $g_{\mQ V\mE}^\text{(meas.)}$ & \\ 
            \hline
            \multirow{4}{*}{$\chi_{c1}(1P)$} & \multirow{4}{*}{$\gamma$} & $\rho$ & -- & 0.019 & -- & \cite{BESIII:2011ysp}\\
            && $\omega$ & -- & 0.010  & -- & \cite{BESIII:2011ysp} \\
            && $\phi$ & -- & $0.005$  & -- & \cite{BESIII:2011ysp}\\
            && $J/\psi$ & -- & 1.29 & -- & 
           ~\cite{ParticleDataGroup:2022pth} \\
            \hline
            \multirow{4}{*}{$X(3872)$} & \multirow{2}{*}{$\gamma$} & $\gamma$ & $3.2\E{-3}$ & -- & -- &~\cite{Belle:2020ndp} \\
            &  & $J/\psi$ & -- & 0.118 & -- &~\cite{Belle:2011wdj} \\
            &\multirow{2}{*}{$J/\psi$} & $\rho$ & -- & -- & 3.24 & \cite{Albaladejo:2020tzt} \\
            && $\omega$ & -- & -- & 7.31 & \cite{Albaladejo:2020tzt} 
            \end{tabular}
        \end{ruledtabular}
    \end{table*}

\subsection{Tests of VMD}
\label{sec:vmd_tests}

With amplitudes for photon and meson exchange in place, we first compare couplings of the $\chi_{c1}(1P)$ and $X(3872)$ determined from the observed radiative decays $\mQ\to \gamma\mE$ with the VMD expectations given by  \cref{eq:VMD_couplings_radiative}.
For radiative decays, the coupling can be computed by evaluating
    \begin{align}
        \label{eq:radiative_width}
        \Gamma_{\mQ\to \gamma\mE} = \, &\left(\frac{m_\mQ^2 - m_\mE^2}{16 \, m_\mQ^3} \right)
        \\
        \times \frac{1}{3} &  \, \sum_{\lambda_\mQ, \,\lambda_\gamma} \, g_{\mQ\gamma\mE}^2 \, \mT^{\mu}_{\lambda_\mQ,\lambda_\gamma} \, (-g_{\mu\nu}) \,\mT^{*\nu}_{\lambda_\mQ,\lambda_\gamma}  ~,
        \nonumber
    \end{align}
where the top vertex is given in \cref{eq:mT}.
The two-photon reduced width is defined by:
    \begin{align}
        \label{eq:reduced_width}     
        \tilde{\Gamma}_{\mQ\to\gamma\gamma} &= \lim_{m_\mE^2\to0} \left(\frac{m_{\mQ}^2}{m_\mE^2}\right) \,\Gamma_{\mQ\to\gamma\mE} ~.
    \end{align}
All couplings extracted from measurements are summarized in \cref{tab:raw_couplings}. We remark that the relative phases between couplings of different exchanges are unknown and are assumed to be real and positive unless otherwise stated.

Starting with the $\chi_{c1}(1P)$, because the radiative decays have all been measured, we can use VMD as in \cref{eq:VMD_couplings_photon} to predict the two-photon coupling: 
    \begin{equation}
        \label{eq:chic1_coupling}
        g_{\chi_{c1}\gamma\gamma} = \sum_{i=\rho,\omega,\phi,\psi} \frac{g_{\chi_{c1}i\gamma}}{\gamma_i} = 3.6 \E{-2} ~.
    \end{equation}
Experimentally, BESIII has observed a signal for the direct production process $e^+e^- \to \chi_{c1} \to J/\psi(\to \mu^+\mu^-) \gamma$ ~\cite{BESIII:2022mtl} but has not extracted the reduced two-photon width yet. We therefore instead compare \cref{eq:chic1_coupling} with other theoretical predictions which rely on specific microscopic models.
In nonrelativistic QCD, the reduced two-photon width can be computed from the $c\bar{c}$ bound-state wave function and predicted to be $\tilde{\Gamma}_{\chi_{c1} \to \gamma\gamma} \sim 0.93\kev$ \cite{Achasov:2022puf}.  
Another calculation based on the light-front wave functions predicts $\tilde{\Gamma}_{\chi_{c1}\to\gamma\gamma} = 3.0\kev$~\cite{Li:2021ejv}. These two values of the width yield couplings of $g_{\chi_{c1}\gamma\gamma} = 0.9 \E{-2}$ and $1.6\E{-2}$, respectively, which are of the same order of magnitude as the VMD prediction \cref{eq:chic1_coupling}. We note that the $J/\psi$ term is dominant in the sum of \cref{eq:chic1_coupling} as the other contributions are OZI-suppressed, and therefore small. 

Turning to the $X(3872)$, there are several ways to compare the known coupling with VMD expectations. Using \cref{eq:VMD_couplings_photon} and ignoring OZI-suppressed terms, the two-photon coupling is related to the radiative decay into $J/\psi$,
    \begin{equation}   
        \label{eq:Xgg}
        g_{X\gamma\gamma} =  \frac{g_{X\psi\gamma}^\text{(meas.)}}{\gamma_\psi} = 3.2 \E{-3}~,
    \end{equation}
which agrees very well with the coupling extracted from the Belle measurement in \cref{tab:raw_couplings}.

For the radiative decays in to light mesons, one may use \cref{eq:VMD_top} to determine them from the known two-hadron couplings. This method was previously employed in Ref.~\cite{Albaladejo:2020tzt} and will be referred to as ``VMD 1". A test of the accuracy of this estimation is to use \cref{eq:VMD_couplings_radiative} to recalculate the $J/\psi\gamma$ radiative coupling from the extracted light meson contributions:
    \begin{equation}
        g_{X\psi\gamma} = \frac{g_{X\psi\rho}^\text{(meas.)}}{\gamma_\rho} + \frac{g_{X\psi\omega}^\text{(meas.)}}{\gamma_\omega} = 0.329 ~.
    \end{equation}
Compared to the measured value in \cref{tab:raw_couplings}, the resulting coupling is within the correct order of magnitude but is deviated by a factor of $\sim3$. 

From this comparison, we see that, within VMD, the size of the extracted two-hadron couplings implies the radiative couplings into light mesons may still be sizeable even after the OZI suppression. Thus we may choose to combine \cref{eq:VMD_couplings_photon,eq:VMD_couplings_radiative} to get an estimate more consistent with the measured couplings. Specifically, because the $X\psi\gamma$ contribution saturates the two-gamma coupling as calculated in \cref{eq:Xgg}, the sum over light vector terms in \cref{eq:VMD_couplings_photon} must vanish.~\footnote{We note that the statistical uncertainties of the measured two-photon width permit $\tilde{\Gamma}_{X\to\gamma\gamma} = 20 - 500$ eV \cite{Belle:2020ndp}. Using the upper bound value results in a factor of two increase in $g_{X\gamma\gamma}$ relative to the one quoted in \cref{tab:raw_couplings}.} This implies that the couplings to the $\rho$ and $\omega$, while comparable to the $J/\psi$ radiative coupling, will interfere destructively. We thus have:
    \begin{equation}
        \label{eq:Xgg_diff}
        \frac{g_{X\gamma\rho}}{\gamma_\rho} + \frac{g_{X\gamma\omega}}{\gamma_\omega} = g_{X\gamma\gamma}^\text{(meas.)} - \frac{g_{X\gamma\psi}^\text{(meas.)}}{\gamma_\psi} = 0 ~,
    \end{equation}
which, combined with \cref{eq:VMD_top},  may be used to determine the radiative couplings by:
    \begin{equation}
        \label{eq:Xgr}
        g_{X\gamma\rho} = -\frac{\gamma_\rho}{\gamma_\omega} \, g_{X\gamma\omega} = - \frac{\gamma_\rho}{\gamma_\omega} \, \frac{g^\text{(meas.)}_{X\psi\omega}}{\gamma_\psi} ~.
    \end{equation}
Here we use the measured $X\psi\omega$ coupling to determine both couplings but we may alternatively invert this relation and use the measured $X\psi\rho$ coupling as input instead:
      \begin{equation}
        \label{eq:Xgo}
        g_{X\gamma\omega} = -\frac{\gamma_\omega}{\gamma_\rho} \, g_{X\gamma\rho} = - \frac{\gamma_\omega}{\gamma_\rho} \, \frac{g^\text{(meas.)}_{X\psi\rho}}{\gamma_\psi} ~.
    \end{equation}  
\Cref{eq:Xgr,eq:Xgo} will result in slightly different coupling estimations which we refer to as ``VMD 2" and ``VMD 3" respectively. 

\begin{figure*}
        \centering
        \includegraphics[width=\textwidth]{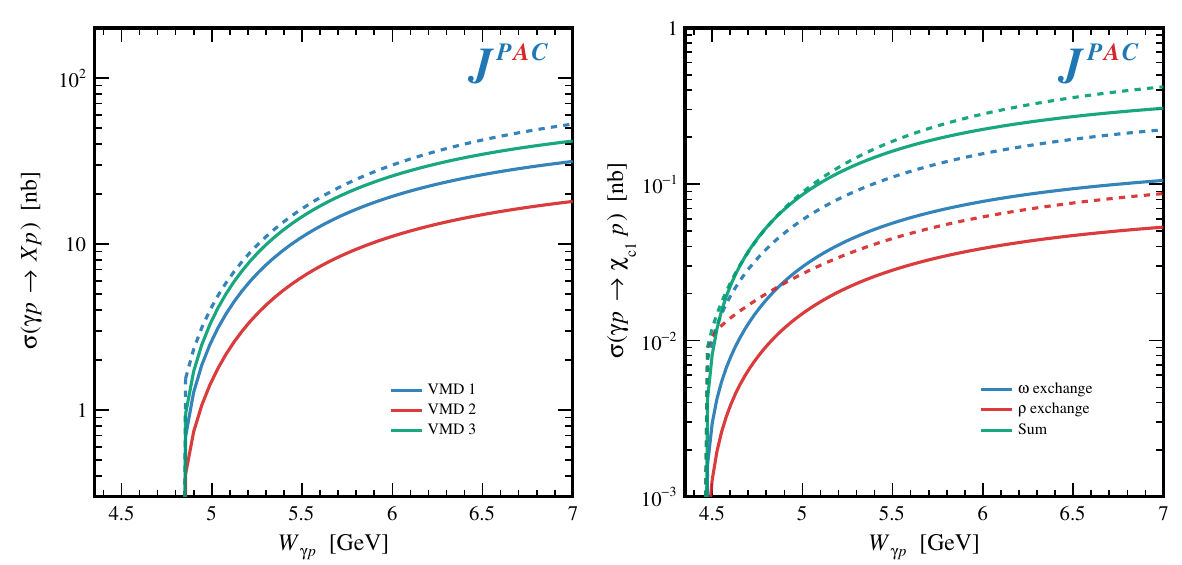}
        \caption{Predictions for exclusive axial-vector meson photoproduction near threshold. In both panels the cross sections are calculated by \cref{eq:vector_exchange} (solid) or by \cref{eq:explicit_meson} (dashed) with the different contributions summed at the amplitude level. Left panel: the $X(3872)$ is calculated with the different estimations of top couplings with VMD as described in the text and in \cref{tab:X_couplings}. Right panel: the $\chi_{c1}(1P)$ is shown with individual meson contributions. }
        \label{fig:exc_compare}
    \end{figure*}

We calculate the top couplings for $\rho$ and $\omega$ exchange using the three VMD methods described in \cref{tab:X_couplings}, where we see a consistent prediction with respect to the absolute size of each contribution. 
The similarity in sizes and, importantly, the effect of the possible relative sign, can be seen in the resulting cross sections in the left panel of \cref{fig:exc_compare}. The spread of predictions is inherent to the model dependence of VMD and should be taken as an estimate of the systematic uncertainty. All together, the predicted cross sections based on VMD are sizeable and not at odds with current measured decay widths. In the numerical studies of semi-inclusive productions we will continue to use the ``VMD 1" couplings as these are consistent with previous predictions in Ref.~\cite{Albaladejo:2020tzt} and provide a more central value for the cross section.

We also note that, while the two-photon coupling of the $X(3872)$ is much smaller than the single radiative ones, the $\chi_{c1}(1P)$ couplings to $\gamma(\rho/\omega)$ and $\gamma\gamma$ are of comparable size.
This will reflect in an inverted hierarchy of the photon and vector meson exchange contributions at high energies between the $\chi_{c1}(1P)$ and the $X(3872)$, as we will show in \cref{sec:numerical}.

\begin{table}[]
    \centering
    \caption{Three different applications of VMD to estimate the charmless radiative couplings of the $X(3872)$ relevant for the top vertex. We assume all couplings to be real and positive except in VMD 2 and 3 where the $\rho$ coupling is chosen negative to saturate \cref{eq:Xgg_diff}.}
    \begin{ruledtabular}
    \begin{tabular}{cccc}
        & $|g_{X\gamma\rho}|$ & $|g_{X\gamma\omega}|$ & Eq. \\
        \hline
        VMD 1 &  $0.088$ & 0.199 & (\ref{eq:VMD_top}) \\
        VMD 2  & $0.058$ & 0.199 & (\ref{eq:Xgr})\\
        VMD 3  & $0.088$ & 0.303 & (\ref{eq:Xgo})
    \end{tabular}
    \end{ruledtabular}
    \label{tab:X_couplings}
\end{table}

Finally, as this will be crucial to the semi-inclusive extension, we use the $\chi_{c1}(1P)$ to test the form of the bottom vertex in \cref{eq:vector_exchange} to reproduce predictions based on meson-nucleon couplings.
We fix the top couplings to the measured values in \cref{tab:raw_couplings} and compute the meson exchange amplitudes using either \cref{eq:Gammaq_meson} or with \cref{eq:Gammaq_photon} rescaled by \cref{eq:VMD_bottom,eq:beta}.
The resulting curves are shown for the $\omega$ and $\rho$ exchanges in the right panel of \cref{fig:exc_compare}, where we see that the bottom vertex in \cref{eq:vector_exchange} reproduces previous predictions of the sum of the two vector meson exchange amplitudes well near threshold. 
We consider this as a validation of the form in \cref{eq:vector_exchange} to relate vector meson and photon exchanges and can use this form to extend the calculation to the semi-inclusive production in the following section.

\section{Semi-inclusive formalism}
\label{sec:SemiInclusive}

Following Ref.~\cite{Winney:2022tky}, the $\mE p \to$ anything transition in the bottom vertex depends on (polarized) vector meson-nucleon inclusive cross sections, which 
are unfortunately not accessible  experimentally. We thus resort to VMD to relate the meson-nucleon vertex to the photoabsorption cross sections of the proton, which are well known over a broad range of kinematics. 

We consider the double differential inclusive cross section via photon exchange, as a function of the momentum transfer $t$ and invariant missing mass $\M$. This process is shown diagrammatically in \cref{fig:momlabels}. We can immediately write it in analogy to the cross section expressions in deep inelastic scattering (DIS) of electrons off nucleons:~\footnote{In DIS contexts our $t$ and $\M$ are typically denoted $-Q^2$ and $W$ respectively. The latter should not be confused for our $W_{\gamma p}$ which is the invariant mass of the \textit{beam}-target system.}
    \begin{equation}
        \label{eq:d2sigma}
        \frac{d^2\sigma_{\text{$\gamma$-ex.}}}{dt \, d\M^2} = \frac{g_{\mQ\gamma\gamma}^2 \, e^2
        }{8\pi\, (2\sqrt{s} \, E_\gamma)^2} \, \mathcal{T}_{\gamma\mQ}^{\mu\nu} \, \left|\mathcal{P}_{\gamma} \right|^2   W_{\mu\nu} ~,
    \end{equation}
with $E_\gamma = (s - m_N^2)/2\sqrt{s}$ the beam energy in the CM frame. Details of kinematics, definitions, conventions, and relations to matrix elements are given in \cref{app:kinematics}. 
Here the propagator with no explicit Lorentz indices refers simply to $\mP_\gamma = 1/t$, where $t = q^2$ is the virtuality of the spacelike exchanged photon.
The lower bound of $\M$ is given by the lowest multiparticle threshold, $M_\text{min} = m_N + m_\pi$. This formula therefore does not account for the exclusive $\mQ \,p$ final state, which has to be added separately.

The bottom tensor is parameterized as (see e.g.~\cite{ParticleDataGroup:2022pth}):
\begin{widetext}
    \begin{align}
        \label{eq:W_def}
        W^{\mu\nu} &= \left(-g^{\mu\nu} + \frac{q^\mu q^\nu}{t} \right) \, F_1 +
        \frac{1}{p\cdot q} \left(p^\mu - \frac{p\cdot q}{t} \, q^\mu\right)\left(p^\nu - \frac{p\cdot q}{t} \, q^\nu\right)\, F_2 ~,
    \end{align}
in terms of the proton structure functions $F_{1,2} \equiv F_{1,2}(\xB,t)$ which we denote as functions of momentum transfer and the Bjorken scaling variable, 
    \begin{equation}
        \label{eq:xb}
        \xB = \frac{-t}{2\,\sp{p}{q}} = \frac{-t}{\M^2 - m_N^2 -t}~.
    \end{equation}
The top tensor comes from the interaction in \cref{eq:mT}:
    \begin{align}
        \label{eq:mT_tensor}
        \mT^{\mu\nu}_{\gamma\mQ} &\equiv \frac{1}{2}\sum_{\lambda_\gamma\lambda_\mQ} \mT^{\mu}_{\lambda_\gamma\lambda_\mQ} \, \mT^{*\nu}_{\lambda_\gamma\lambda_\mQ} 
        = \left(-g^{\mu\nu} + \frac{q^\mu \, q^\nu}{t} \right) \, T_1 + \frac{1}{k\cdot q} \left(k^\mu - \frac{k\cdot q}{t} \,q^\mu\right)\left(k^\nu - \frac{k\cdot q}{t} \,q^\nu\right) \, T_2
        ~,
    \end{align}
\end{widetext}
where  
\begin{subequations}
    \label{eq:Ts}
    \begin{align}
        T_1 & \equiv  |f_{\mQ\gamma\gamma}|^2 \, \frac{t^2}{2\,m_\mQ^6} (k\cdot q)^2 ~, \\
        T_2 & \equiv  |f_{\mQ\gamma\gamma}|^2 \,  \frac{t^2}{2\, m_\mQ^6} \, (k\cdot q) \, \left[m_\mQ^2 - 2 \, (k\cdot q) \right] ~,
    \end{align}
\end{subequations}
and after performing  Lorentz contractions,
    \begin{align}
        \label{eq:TdotW}
        \mT^{\mu\nu}_{\gamma \mQ} &\; W_{\mu\nu} = 
        3\, F_1 \, T_1
        \nonumber \\
        &+  \left[\frac{\sp{k}{q}}{t} \right] F_1 \, T_2 + \left[\frac{\sp{p}{q}}{t} - \frac{m_N^2}{\sp{p}{q}}\right] F_2 \, T_1 
         \\
        &+ \left[ \frac{\sp{k}{q}\sp{p}{q}}{t^2} - 2 \,\frac{\sp{p}{k}}{t} + \frac{\sp{p}{k}^2}{\sp{p}{q}\sp{k}{q}} \right]\, F_2 \, T_2 ~.
        \nonumber
    \end{align}

Using the same argument that led to \cref{eq:vector_exchange}, to obtain  the vector exchange formula from \cref{eq:d2sigma}, we need to first replace the electromagnetic couplings with the hadronic ones calculated with VMD, and then replace the massless propagator with a sum over the massive ones, \ie
\begin{equation}
    \label{eq:VMD_replacement}
    g_{\mQ\gamma\gamma} \, \mP_\gamma \, e \to \sum_{\mE = \rho,\omega} \,  g_{\mQ\gamma\mE} \, \mathcal{P}_{\mE} \, \beta_\mE  \, \frac{\gamma_\mE}{2}\,  e ~,
\end{equation}
with the $\gamma_\mE$ factors of \cref{eq:VMD_couplings} and form factor ratio $\beta_\mE$ of \cref{eq:beta}. In $\beta_\mE$ we continue to use \mbox{$t' = t - t(s, \cos\theta = 1, \M^2 = m_N^2)$} to match the form factors in the exclusive reaction and not introduce spurious $\M$ dependence. 

As discussed in \cref{sec:vmd_tests}, the sum is restricted to $\rho$ and $\omega$, as the interaction between the proton and the other vectors is OZI-suppressed just as in the exclusive case. Since the left-hand side of \cref{eq:VMD_replacement} is squared in \cref{eq:d2sigma}, the inclusive cross section contains a double sum which includes the $\rho N \to \omega N$ nondiagonal transition.
The resulting cross section via meson exchanges is thus:
    \begin{align}
        \label{eq:d2sigma_mesons}
        &\frac{d^2\sigma_{\text{$\mE$-ex.}}}{dt \, d\M^2} =  \frac{e^2}{128 \pi \, s \, E_\gamma^2} 
        \\ 
        &\quad\times \mathcal{T}_{\gamma\mQ}^{\mu\nu} \,  |g_{\mQ\gamma\rho} \,  \mathcal{P}_{\rho} \, \gamma_\rho \, \beta_\rho + g_{\mQ\gamma\omega} \,  \mathcal{P}_{\omega} \, \gamma_\omega \, \beta_\omega|^2  \,  W_{\mu\nu} ~.
        \nonumber
    \end{align}
This contribution can be added to the photon exchange cross section in \cref{eq:d2sigma} to estimate the combined production. We note that we add the two terms incoherently and ignore the interference between strong and electromagnetic exchanges for simplicity. In the energy regions of interest, one term will be found to dominate by several orders of magnitude and this interference is negligible.
\Cref{eq:d2sigma_mesons} is valid at energies where a description in terms of fixed-spin exchanges is applicable, and we restrict this form to energies within $\sim2$\gev from threshold. At higher energies, to avoid violations of unitarity, higher-spin exchanges must be included and resummed into Reggeized meson propagators. 

In the exclusive case, Reggeization of a spin-$j$ exchange can be implemented with a prescription that replaces the leading powers of $s^j$ in the propagator with $(s/s_0)^{\alpha(t)}$, where $\alpha(t)$ is the exchanged Regge trajectory and $s_0 = 1\gevsq$ is a characteristic hadronic scale (c.f. Eq.~6 in Ref.~\cite{Albaladejo:2020tzt}). In semi-inclusive production, matching the leading $s$ behavior to the expected behavior of the inclusive production distribution in the triple Regge limit ($-t \ll \M^2 \ll s$) motivates the form of the Regge propagator with respect to the missing mass $\M$ instead of $s_0$:
    \begin{align}
        \label{eq:regge_perscription}
        \left|\frac{1}{t-m_\mE^2}\right|^2& \left(\frac{s}{\M^2}\right)^{2} \to
        \\
        &\left|\alphaV^\prime \, \xi(t) \, \Gamma(1 - \alphaV(t))\right|^2 \, \left(\frac{s}{\M^2}\right)^{2 \,\alphaV(t)} ~.
         \nonumber
    \end{align}
Here $\alphaV(t) = \alphaV(0) + \alphaV^\prime \, t$, with $\alphaV(0)=0.5$ and \mbox{$\alphaV^\prime =0.9\gev^{-2}$} for 
both $\rho$ and $\omega$. The signature factor is given by:
    \begin{equation}
        \xi(t) = \frac{1}{2} \left[-1 +  e^{-i\pi \, \alphaV(t)} \right]~.
    \end{equation}

In \cref{eq:d2sigma_mesons}, the $s$ dependence may only come from the kinematic prefactors of the two hadronic tensors and, from \cref{eq:TdotW}, we can identify the form
    \begin{equation}
        \label{eq:gamma_Expand}
        \mT^{\mu\nu}_{\gamma\mQ} \, \mathcal{P}^2_\mE \, W_{\mu\nu} = \sum_{n =0}^{2} \, C_n \,\left|\frac{1}{t-m_\mE^2}\right|^2 \, \left(\frac{s}{\M^2}\right)^{2-n} ~,
    \end{equation}
where the coefficients are functions of $t$ and $\M$:
\begin{subequations}
    \label{eq:gammas}
    \begin{align}
        C_0 &= \frac{\M^4}{4 \, \sp{p}{q} \, \sp{k}{q}} \, T_2 \, F_2 ~, \\
        C_1 &= - \left(\frac{\M^2}{t} {+\frac{\M^2 \, m_N^2}{2(k\cdot q)(p \cdot  q)}}\right) T_2 \, F_2 ~, \\
        C_2 &= 3 \, F_1 \, T_1  + \frac{\sp{k}{q}}{t} \, F_1 \, T_2 + \left[\frac{\sp{p}{q}}{t} - \frac{m_N^2}{\sp{p}{q}}\right] F_2 \, T_1 
        \nonumber 
        \\
        + &\left[\frac{\sp{k}{q} \sp{p}{q}}{t^2} { +\frac{m_N^4}{4(k\cdot q)(p\cdot q)} +\frac{m_N^2}{t}}\right] F_2 \, T_2 ~.
    \end{align}
\end{subequations}
We have three terms because of the different subleading powers of $s$ which appear from the half-angle factors when exchanging a particle with spin (e.g. $\sin\theta/2 \to \sqrt{-t/s}$ at fixed $t$ and large $s$). Clearly, as $s\to \infty$, only the $n=0$ term will survive, but we may keep all terms to capture some subleading behavior.
Because the $C_n$ functions are independent of $s$, the Reggeization prescription of \cref{eq:regge_perscription} can be applied directly, yielding
\begin{widetext}
    \begin{align}
        \label{eq:d2sigma_regge}
        &\frac{d^2\sigma_{\text{$\mathbb{R}$-ex.}}}{dt \, d\M^2} = \frac{e^2 
        }{128 \pi \, s \, E_\gamma^2} \, |g_{\mQ\gamma\rho} \, \gamma_\rho \, \beta_\rho + g_{\mQ\gamma\omega} \,  \gamma_\omega \, \beta_\omega|^2   \, \sum_{n = 0}^2 \, C_n \, |\alphaV^\prime \, \xi \, \Gamma(1-\alphaV(t))|^2 \, \left(\frac{s}{\M^2}\right)^{2\,\alphaV(t) - n} ~. 
    \end{align}
\end{widetext}
At high energies, this expression can be added to the photon exchange which does not Reggeize~\cite{Gribov:2009zz}.  

\subsection{Proton structure functions}
\label{sec:NT}
For the structure functions $F_{1,2}$, we need to separately consider kinematic regimes which represent different physics in the bottom vertex. For production close to threshold, which is expected to have the largest cross sections, the structure functions are evaluated at missing masses of only a few\gev, and are dominated by nucleon resonances.  

\begin{figure}[t]
        \centering
        \includegraphics[width=\columnwidth]{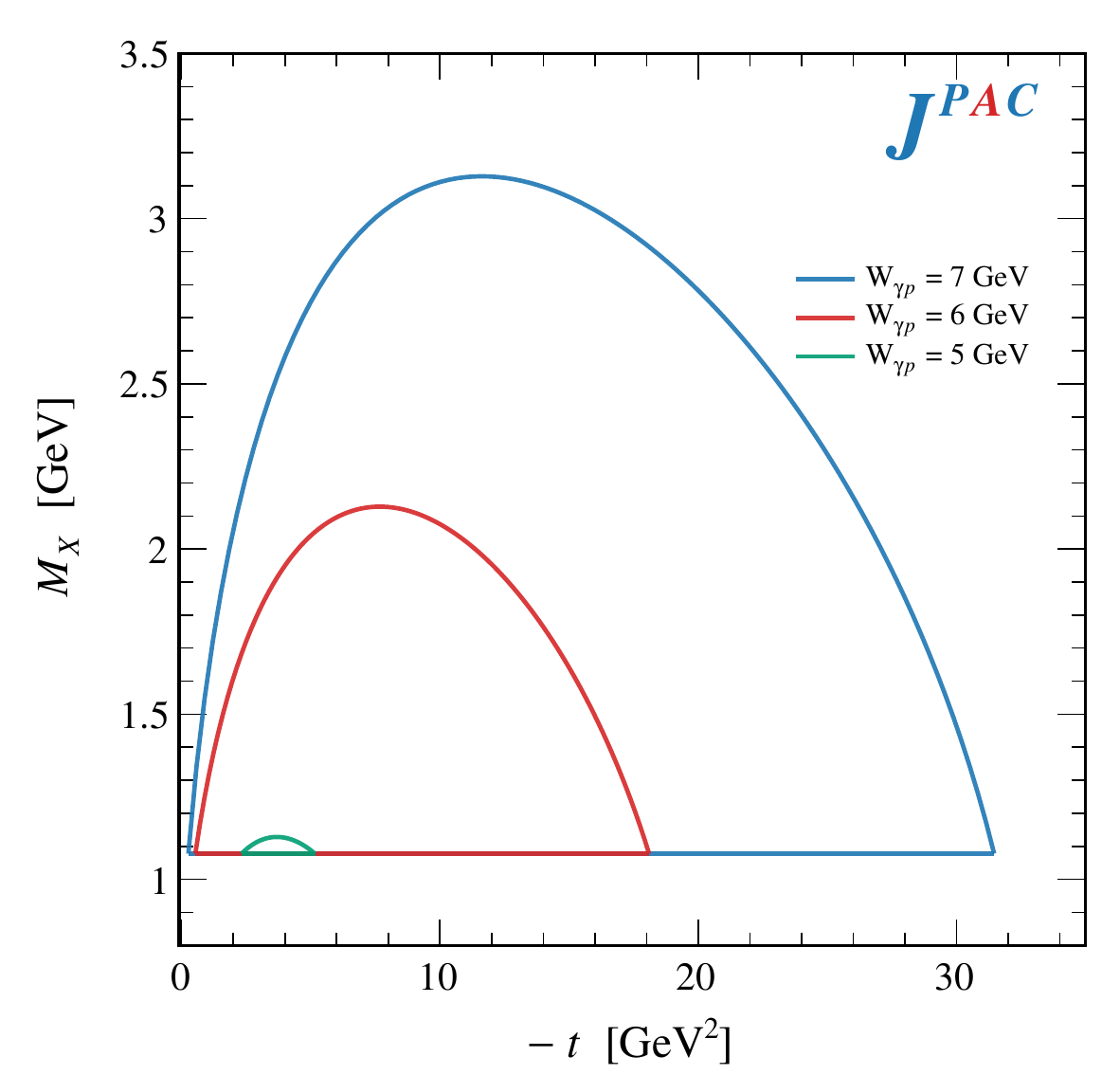}
        \caption{Chew-Low plot for inclusive $X(3872)$ production at near threshold energies. The lower bound of missing mass is given by the lowest inelastic threshold, $M_{\text{min}} = m_N + m_\pi$.}
        \label{fig:chew_low}
    \end{figure}

For the near-threshold region, we use the parameterizations of Refs.~\cite{Bosted:2007xd,Christy:2007ve} (herein referred to as B\&C), which describe current electron-proton inclusive data in the resonance region ($\M \leq 3.5\gev$ and $-t \le 7.5\gevsq$). 
We note that this parameterization is an empirical description of the data, thus extrapolating far outside of the fitted region may yield unphysical behavior.
To consider energies $W_{\gamma p} \leq 7\gev$, we must evaluate the structure functions for $\M \lesssim 3\gev$ and $0\lesssim -t \lesssim 30\gevsq$ as illustrated in the Chew-Low plots in \cref{fig:chew_low} and we thus need to extrapolate to large momentum transfers. Luckily both $F_{1,2}$ vanish quickly as $|t|\to \infty$ (at fixed $\M$ this implies $\xB \to 1$) and we are not very sensitive to details of the poorly constrained high-$t$ region.

When increasing $W_{\gamma p}$, higher values of {\M}  become accessible.
Extrapolation of B\&C to higher $\M$ however, is not well-behaved and considering center-of-mass energies above $W_{\gamma p} \gtrsim 7\gev$ requires a different parametrization. 
    \begin{figure*}[t]
        \centering
        \includegraphics[width=\textwidth]{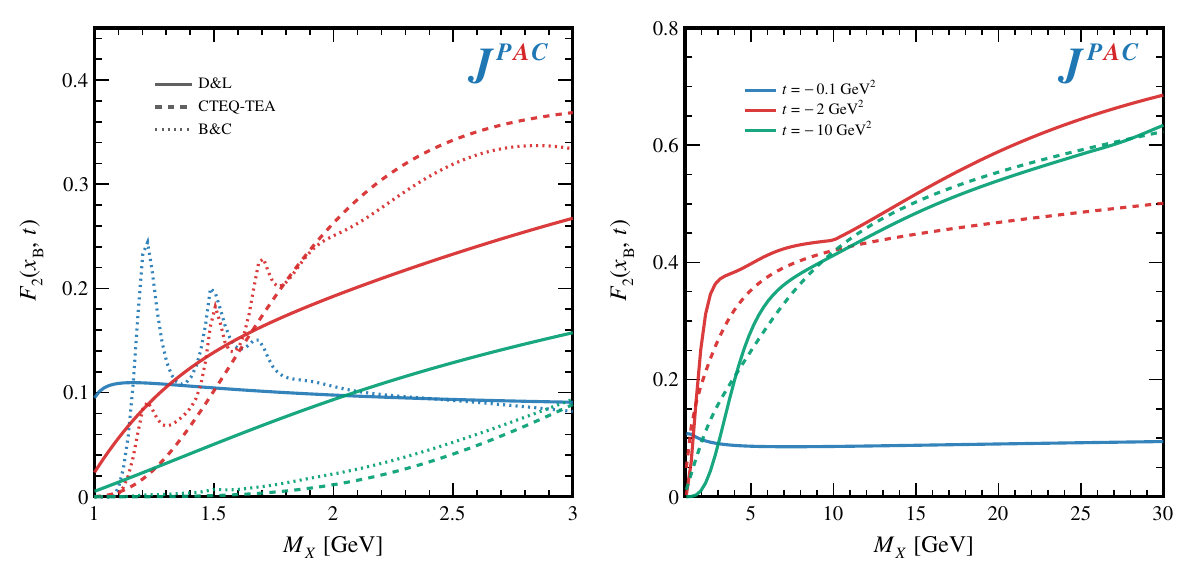}        
        \caption{Comparisons of the structure function $F_2$ using the D\&L parameterization \cref{eq:F2_DL} (solid), the CTEQ-TEA's LO PDFs (dashed) and B\&C (dotted). These are evaluated at fixed $t$ in the near-threshold region (left panel), as well as at large missing masses (right panel). Three representative values of $t$ are shown to exemplify the $F_{2} \to 0$ as $-t \gg \M^2$ (i.e. $\xB \to 1$) behavior at small missing mass.
        } 
        \label{fig:DL_Fs}
    \end{figure*}
At these high energies, the peaks of the resonance region make up a small portion of the entire phase space, and the details of their descriptions are not relevant. We thus use a phenomenological Regge-inspired parameterization from Refs.~\cite{Donnachie:2004pi,Tomalak:2015hva} (referred to as D\&L):
    \begin{equation}
        \label{eq:F2_DL}
        F_{2}(\xB,t) = \sum_{k} \, f_{2}^k(t) \, \left(1- \xB\right)^{\epsilon_k} \, \xB^{1 - \alpha_k(0)} ~.
    \end{equation}
Here the sum runs over two Pomerons (hard and soft~\cite{Donnachie:1998gm}) and a Reggeon. 
This form is based on Regge factorization arguments at small $\xB$, and thus additional powers of $(1-\xB)$ are included to enforce that the structure functions vanish at $\xB = 1$. Based on quark counting rules, one takes $\epsilon_k = 5$ for Pomeron exchanges and $\epsilon_k=1$ for Reggeons. These factors are simplistic and do not account for realistic QCD evolution, but enforce a slightly more physical behavior in regions with large $t$ at small $\M$. At any rate, the production is dominated by the forward limit where $\xB \sim 0$, and predictions are relatively insensitive to the exact values of $\epsilon_k$.

The Regge pole residues are given by the form:
    \begin{equation}
        \label{eq:f2_DL}
        f_{2}^k(t) = A_k \, \left(\frac{-t}{s_0}\right)^{\alpha_k(0)} \left[1 + \frac{- t}{\Lambda_k^2}\right]^{n_k - \alpha_k(0)} ~,
    \end{equation}
with intercept $\alpha(0)$, normalization $A$, and scale $\Lambda^2$ for each exchange given in \cref{tab:DL_F2pars}. The exponent factor $n_k$ adjusts the $t$-dependence of the coupling only for the hard Pomeron and is defined as $n_{\mathbb{P}_\text{hard}} = (\alpha_{\mathbb{P}_\text{hard}}(0) - 1)/2$ while $n_k =0$ for the other exchanges.
        \begin{table}[h]
            \centering
            \caption{Summary of parameters for the D\&L Regge pole parameterization~\cite{Donnachie:2004pi} for the $F_2$ structure function.}
            \begin{tabular}{cccc}
            \hline\hline
                $k$ & $A$ & $\alpha(0)$ & $\Lambda^2$ [$\gevp^2$] \\
            \hline
                $\mathbb{P}_\text{hard}$ & 0.00151 & 1.452  & 7.85 \\
                $\mathbb{P}_\text{soft}$ & 0.658   & 1.0667 & 0.6 \\
                $\mathbb{R}$ & 1.01    & 0.524 & 0.214
                 \\
            \hline \hline
            \end{tabular}
            \label{tab:DL_F2pars}
        \end{table}

The structure function $F_1$ can be calculated from $F_2$ using the ratio of longitudinal-to-transverse photoabsorption cross sections as parameterized in Ref.~\cite{Tomalak:2015hva}. 

As a comparison point, we compute the same structure functions using parton distribution functions (PDFs) extracted from global fits. At leading order, we may calculate:
    \begin{equation}
        \label{eq:CTEQ_F2}
        F_2(\xB, t) = \sum_{i} e_i^2 \, \xB\, \phi_i(\xB, t) ~,
    \end{equation}
where we sum over all parton flavors of charge $e_i$ and with PDF $\phi_i$. 
The other structure function may be constructed through the Callan-Gross relation.
We interpolate the most recent leading order PDFs from the \mbox{CTEQ-TEA} Collaboration~\cite{Yan:2022pzl}.

The structure function $F_2$, which dominates at large $s$, is plotted in \cref{fig:DL_Fs} for the different parameterizations. 
At small missing masses, the CTEQ and D\&L structure functions are generally compatible with B\&C, with the former two basically averaging through the resonance peaks of the latter. Unfortunately, the largest contribution to the semi-inclusive cross section comes from the small-$t$ region, where the perturbative QCD parameterizations break down. We therefore use B\&C at low $W_{\gamma p}$ and D\&L at high $W_{\gamma p}$ in all phase space, instead of interpolating between the different models.

\subsection{Triple Regge behavior}
\label{sec:TripleRegge}
    \begin{figure*}
        \centering
        \includegraphics[width=\textwidth]{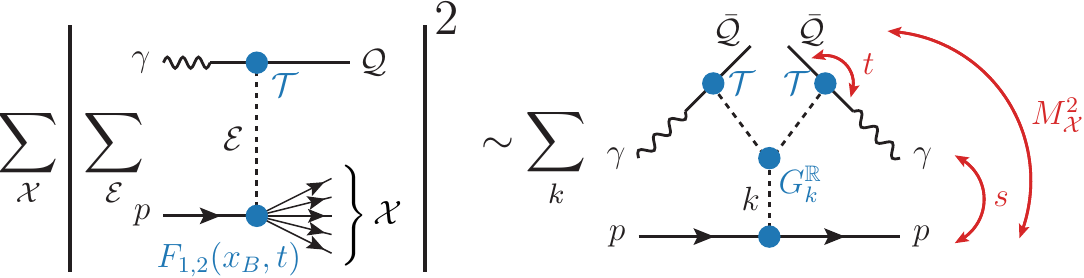}
        \caption{Diagrammatic representation of the triple Regge formula of \cref{eq:tripleregge}. Since the sum over \mE~runs over natural vectors only, the propagators factor out and the sum can be included in the coupling $G_k^\mathbb{R}$. Alternatively, photon propagators and the $G_k^\gamma$ coupling can be used. The sum over $k$ includes all the bottom exchanges in \cref{tab:DL_F2pars} related to the D\&L parameterization.}
        \label{fig:triple}
    \end{figure*}
As done with pion exchange in Ref.~\cite{Winney:2022tky}, it is illustrative to match the cross section to the triple Regge limit (see e.g.~\cite{Field:1974fg}) to examine the expected behavior of the inclusive vector exchange cross sections at high energies. 
Moreover, since the photon exchange does not reggeize~\cite{Gribov:2009zz} and, as we will see, it actually dominates the $\chi_{c1}(1P)$ production, it is important to ensure that the high-energy predictions do not violate the Froissart unitarity bound.  

Taking the $s \to \infty$ limit with large $\M^2$ and small $-t$, we can replace $\xB \sim -t / \M^2$ into \cref{eq:d2sigma_regge} and rewrite the cross section in the form: 
\begin{widetext}
    \begin{align}
        \label{eq:tripleregge}
        E_\mQ \, \frac{d^3\sigma}{d^3k^\prime} &= \frac{(2\sqrt{s} \, E_\gamma)}{\pi} \, \frac{d^2\sigma}{dt\, d\M^2} 
         \equiv \sum_{\,k} \, \frac{1}{\pi\, s_0\, s}\left[ G^\gamma_k(t) \,\left(\frac{s}{\M^2}\right)^{2} +  \Gvvk(t) \,\left(\frac{s}{\M^2}\right)^{2\,\alphaV(t)}  \right] \left(\frac{\M^2}{s_0}\right)^{\alpha_k(0)} ~,
    \end{align}
where the functions $G^\gamma_{k}(t)$ and $\Gvvk(t)$ are the ``triple Regge couplings" for the photon and meson exchange reactions respectively, and the sum over $k$ includes all the bottom exchanges in \cref{tab:DL_F2pars}. This is schematically represented in \cref{fig:triple}.
As before, we sum explicitly the strong and electromagnetic contributions while ignoring the cross term. While the vector meson exchange is replaced by the Regge trajectory, the photon carries fixed spin $j=1$ at all energies, which results in very different asymptotic behavior between the two terms. 

To examine the high-energy behavior of the integrated cross section, we express the cross section \cref{eq:tripleregge} in terms of the fraction of longitudinal momentum of $\mQ$ with respect to its maximum possible value, $\xF = |\vec{k}^\prime_\parallel| / |\vec{k}^\prime_\text{max}|$, where $|\vec{k}^\prime_\text{max}| = \lambda^{1/2}\!\left(s, \mQ^2, M_\text{min}^2\right)/2\sqrt{s}$. In the triple Regge limit,  $\xF \sim 1 - \M^2/s \simeq 1$. For 
the Reggeon exchange, this means approximating
    \begin{align}
        \frac{d^2\sigma_{\text{$\mathbb{R}$-ex.}}}{dt\,d\xF} \simeq \pi \, E_\mQ \,\frac{d^3\sigma_{\text{$\mathbb{R}$-ex.}}}{d^3k^\prime} 
        = \sum_k\frac{\Gvvk(t)}{s_0^2}  (1-\xF)^{\alpha_k(0) -2\,\alphaV(t)} \left(\frac{s}{s_0}\right)^{\alpha_k(0) -1} ~.
    \end{align}
We can read off the triple Regge coupling by collecting all $s$ and $\xF$ independent factors in \cref{eq:d2sigma_regge}:
    \begin{align}
        \label{eq:GVVK}
        \Gvvk(t) =  \frac{(s_0 \, \alphaV^{\prime} \, e)^2}{128 \, \pi} \,  \frac{t^2 \, (2 \, m_\mQ^2 - t)}{m_\mQ^6}\, |f_{\mQ\gamma\gamma}|^2\, f_2^k(t)
        \times |g_{\mQ\gamma\rho} 
        \, \gamma_\rho \, \beta_\rho + g_{\mQ\gamma\omega} \,\gamma_\omega \, \beta_\omega|^2 \,  \big| \xi(t) \, \Gamma(1-\alphaV(t)) \big|^2  ~.
    \end{align}
\end{widetext}
The computation of the integrated cross section therefore involves integrating over both $t_\text{min} \geq t \geq t_\text{max}$ and $0\leq \xF \leq 1$, where the bounds of $t$ integration may introduce total energy dependence as the 
boundary of the Chew-Low plot grows with $s$. We note that such integrals evaluate the cross section in regions of phase space where the triple Regge limit is not realized and the model is less reliable. We must therefore ensure that this region does not contribute significantly to our predictions. Performing the $\xF$ integral one finds 
    \begin{align}
        \label{eq:dsigdt}
        \frac{d\sigma_\text{$\mathbb{R}$-ex.}}{dt} = \sum_k \, \left[\frac{\Gvvk(t)}{s^2_0 \, \delta_k(t)} \right] \, \left(\frac{s}{s_0}\right)^{\alpha_k(0) -1}~.
    \end{align}
 where  $\delta_k(t) \equiv  1 + \alpha_k(0) - 2 \, \alphaV(t)$. 
Because $\alphaV(t) \leq 0.5$ in the physical region where $t < 0$, the integral over $t$ avoids the divergence at $t=0$, since $\delta_k(t) \gtrsim 0.5$
for all bottom exchanges. 
The meson-nucleon form factors make $\Gvvk(t)$ vanish exponentially for growing $-t$. The relevant phase space is thus effectively cut at $t \simeq -2\gevsq$, far from the kinematic limit $t_\text{max} = t(s, \cos\theta = -1)$.~\footnote{The unphysical region where the $\Gamma$ function in \cref{eq:GVVK} overcomes the exponential form factor is explicitly removed via a cutoff in $t$, as done in Ref.~\cite{Albaladejo:2020tzt,Winney:2022tky}.} Therefore, the boundaries in $t$ do not carry additional energy dependence to the fully integrated cross section, which is given by
    \begin{equation}
        \label{eq:triple_integrated}
        \sigma_\text{$\mathbb{R}$-ex.}(\gamma p \to \mQ \mathcal{X}) \sim \sum_k \,  \,\left(\frac{s}{s_0}\right)^{\alpha_k(0)-1} ~.
    \end{equation}
Among the possible bottom exchanges, the one that dominates asymptotically is the soft Pomeron, which gives $\alpha_{\mathbb{P}_\text{soft}}(0) \sim 1$ and produces an almost constant asymptotic cross section.~\footnote{The value $\alpha_{\mathbb{P}_\text{soft}}(0) = 1.07 > 1$ in D\&L mimics the logarithmic growth allowed by the Froissart bound. We note that D\&L's hard Pomeron trajectory leads to $\sqrt{s}$ behavior, but is orders of magnitude smaller than the other exchanges and does not contribute appreciably in the energy range considered.} As this statement does not depend on the top trajectory $\alphaV(t)$, the analysis holds true for any Reggeon, and was seen to hold also for pion exchange~\cite{Winney:2022tky}.

In the case of photon exchange, the analogous coupling of \cref{eq:GVVK}  has a much slower fall off at large virtualities. In particular, one finds 
    \begin{equation}
        G^{\gamma}_{k}(t) =  \frac{(s_0^2 \, e \,g_{\mQ\gamma\gamma})^2}{32 \, \pi} \,  \frac{(2 \, m_\mQ^2 - t)}{m_\mQ^6}\, |f_{\mQ\gamma\gamma}|^2\, f_2^k(t) ~,
    \end{equation}
which from \cref{eq:Fqgg,eq:f2_DL} behaves as $1/t$ at large $t$. 
The integrated cross section $\sigma_{\text{$\gamma$-ex.}}$ may thus pick up an slow energy dependence through  \mbox{$\sigma_{\text{$\gamma$-ex.}} \sim \log|t_\text{max}| \sim \log s$.} In this way the $\gamma\gamma\mathbb{P}$ coupling will be similar to a triple Pomeron coupling~\cite{Field:1974fg}, albeit with the overall size given by electromagnetic couplings.

\section{Numerical results}
\label{sec:numerical}

\begin{figure}
        \centering
        \includegraphics[width=\columnwidth]{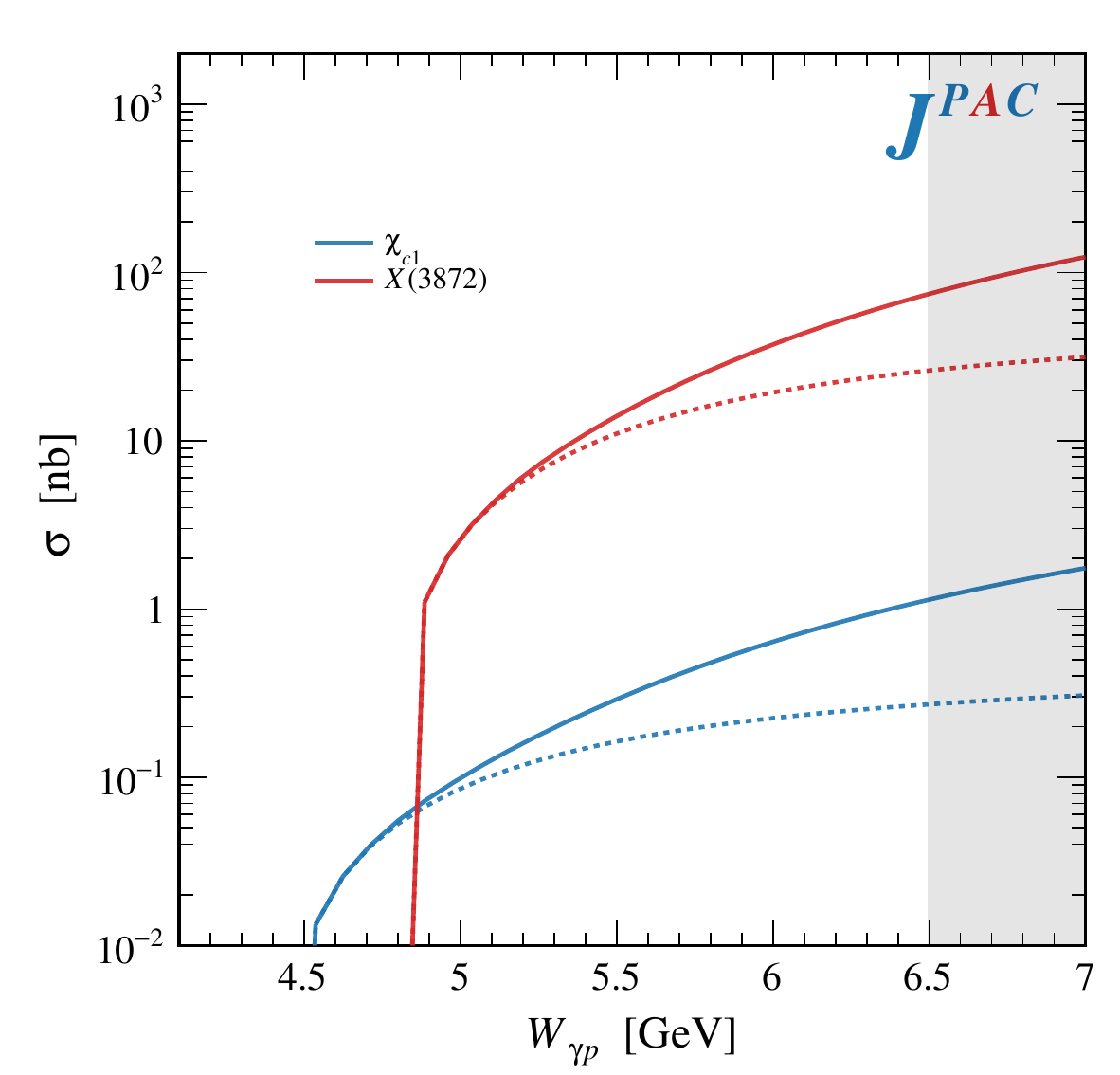}
        \caption{Total integrated photoproduction cross sections for the $X(3872)$ and $\chi_{c1}(1P)$ near threshold. The dashed curves show the contributions from the exclusive final state. The unshaded region marks energies accessible at the proposed JLab upgraded facility with a 22\gev photon beam~\cite{Accardi:2023chb}.}
        \label{fig:NT_X}
    \end{figure}

With all this in place, we may compute the semi-inclusive cross sections with the appropriate parameterizations of the proton structure functions for the kinematic regions of interest for both the $\chi_{c1}(1P)$ and $X(3872)$ production. At energies near threshold, the dominant inclusive contribution at energies near threshold is found to be the isoscalar $\omega$ exchanges, while the contributions of photon exchange is entirely negligible (only a few fb or less) for both cases. The total integrated cross sections for $W_{\gamma }$ near threshold are plotted in \cref{fig:NT_X}. 

The region up to $500\mev$ from threshold is clearly dominated by the $\mQ\,p$ final state, as the available range of missing mass is small. Above it,  other inelastic contributions are sizeable and grow faster with energy than the exclusive final state alone.
As discussed in \cref{sec:SemiInclusive}, the fixed spin-1 exchange models grow indefinitely with $s$, and unitarity requires they must turn downward as Reggeization of the exchange mesons takes place. We do not know \textit{a priori} at what energies this should happen, and therefore the predictions for intermediate energies are uncertain. Still, if one believes that the fixed-spin description is realistic up to $\sim 1\text{--}2\gev$ above threshold, we can already see nearly a factor of $\sim 2$ enhancement over the exclusive reaction. 

Unlike for the pion exchange case and the $\Delta^{++}$~\cite{Winney:2022tky}, no specific nucleon resonance dominates the $\gamma p$ spectrum. We therefore do not isolate any quasi-elastic exclusive processes (e.g. $\gamma p \to X(3872) \Delta^+ (\to p \pi)$), since any individual channel will likely be much smaller than the exclusive reaction. In any case, an experiment with sufficiently high luminosity to observe such channels could provide complementary information to better identify production mechanisms. More physically motivated parameterizations of the structure functions in the resonance region such as in Ref.~\cite{HillerBlin:2019jgp,Blin:2021twt,HillerBlin:2022ltm} can be used in such an analysis.

\begin{figure*}
        \centering
        \includegraphics[width=\textwidth]{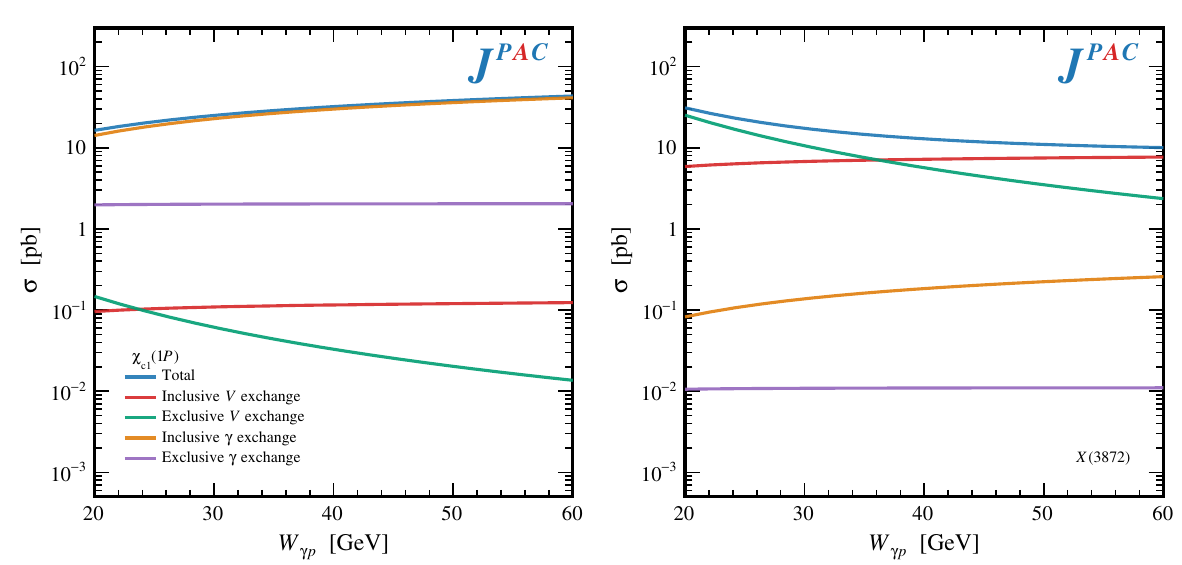}
        \caption{Predicted contributions to the integrated cross section for semi-inclusive $\chi_{c1}(1P)$ (left) and $X(3872)$ (right) production at high energies.}
        \label{fig:HE}
    \end{figure*}

For higher center-of-mass energies, we plot the predicted Reggeized production cross sections in \cref{fig:HE}. We show $W_{\gamma p} \geq 20 \gev$ to avoid the uncertain intermediate energy region. The $\chi_{c1}(1P)$ and $X(3872)$ are found to have different hierarchies with regards to individual production mechanisms: the $\chi_{c1}(1P)$ production is dominated by photon exchange, while the $X(3872)$ by vector meson exchanges. This reflects the different hierarchy of couplings discussed in \cref{sec:vmd_tests} and mirrors the predictions of~\cite{Jia:2022oyl,Benic:2024pqe} which suggest searches for $\chi_{c1}(1P)$ photoproduction at high energies may be dominated by photon exchange. 
\section{Summary and Conclusions}
\label{sec:Conclusions}
In this work, we have generalized previous studies on semi-inclusive photoproduction of (exotic) axial charmonium states, to the case where vector meson exchanges are involved. In particular, we focused on the $\chi_{c1}(1P)$ and $X(3872)$ production. The model is based on the factorization between a ``top vertex'' where the beam dissociates into the observed charmonium and a vector meson, and a ``bottom vertex'' where the vector meson scatters onto the nucleon target. Since the latter interaction is not accessible experimentally, one needs to resort to vector meson dominance to infer it from the measured photon-nucleon cross section. 

The validity of VMD applied to charmonium photoproduction has recently been called into question, as the new data from the GlueX and $J/\psi$-007 experiments~\cite{GlueX:2023pev,Duran:2022xag} suggest a violation of several orders of magnitude~\cite{JointPhysicsAnalysisCenter:2023qgg}.
A sanity check on the applicability of VMD to this type of reactions is therefore opportune. We started by revisiting our exclusive model of~\cite{Albaladejo:2020tzt} and compared the predictions obtained with this model including hadronic vector exchanges with the predictions one would get using a simple VMD rescaling. We found a reasonable consistency between the two within the expected model uncertainties.

With the apparent success of VMD in this sector, we derived a model for semi-inclusive production. We used different parameterizations of the proton structure functions available in the literature, as input to obtain the cross sections of photon and vector meson exchanges. 
We predict the production rates both for energies near threshold, where nucleon resonances must be taken into account, as well as for high energies, where Regge physics must be considered. The results suggest the near-threshold regime has the highest cross sections on the order of several tens of nanobarn and therefore the highest likelihood for observation. Preliminary results from GlueX suggest evidence for $\chi_{c1}(1P)$ production, which will allow us to benchmark our model~\cite{lubomir}.
Furthermore, despite having less constrained kinematics, the semi-inclusive final states were found to be roughly a factor of two enhancement compared to the exclusive final state alone and therefore a promising way to obtain a first observation of the $X(3872)$ in photoproduction at future facilities. We highlight the potential reach of an upgraded JLab facility with a 22\gev photon beam~\cite{Accardi:2023chb} marked by the unshaded region of \cref{fig:NT_X}. 

Cross section predictions at high energies apply to a relatively small portion of the total phase space: where the charmonium state carries most of the beam energy. However, other production mechanisms will also contribute at more central kinematics, so the predictions presented here should be understood as a conservative lower bound of the full semi-inclusive production. Extending our calculation to larger invariant masses, we found substantially smaller production rates, but the triple Regge asymptotics means the inclusive cross section becomes nearly largely independent of $s$. This means that facilities optimized at much higher energies such as the EIC~\cite{Burkert:2022hjz,AbdulKhalek:2021gbh} or the EicC~\cite{Anderle:2021wcy} may also be able to observe these states in inclusive searches.  
The curves shown should be considered as order-of-magnitude estimates, due to the unknown matching between the low and high-energy predictions and the inherent model dependence of VMD.

Code to reproduce all figures and results are available online~\cite{zenodo}.

\acknowledgments
This work was supported by the U.S. Department of Energy contract DE-AC05-06OR23177, under which Jefferson Science Associates, LLC operates Jefferson Lab
and also by the U.S.~Department of Energy Grant 
Nos.~DE-FG02-87ER40365. The work of DW was supported in part  by DFG and NSFC through funds provided to the Sino-German CRC 110 “Symmetries
and the Emergence of Structure in QCD” (NSFC Grant No. 12070131001,
DFG Project-ID 196253076).
VM and RJP have been supported by the projects CEX2019-000918-M (Unidad de Excelencia “María de Maeztu”), PID2020-118758GB-I00, financed by MICIU/AEI/10.13039/501100011033/ and FEDER, UE, as well as by the EU STRONG-2020 project, under the program H2020-INFRAIA-2018-1 Grant Agreement No. 824093.
CFR is supported by Spanish Ministerio de Ciencia, Innovación y Universidades \mbox{(MICIU)} under Grant No.~BG20/00133.
VM is a Serra Húnter fellow and acknowledges support from CNS2022-136085.
NH is supported by Project No. 2018/29/B/ST2/02576, financed by the National Science Center in Poland. VM is a Serra H\'unter fellow. 
VS and WAS acknowledge the support of the U.S.~Department of Energy ExoHad Topical Collaboration, contract DE-SC0023598.
This work contributes to the aims of the U.S.~Department of Energy ExoHad Topical Collaboration, contract DE-SC0023598.

\appendix
\section{Comparison with DIS formulae} 
\label{app:kinematics}
In this appendix, we provide a derivation of \cref{eq:d2sigma} and relate to the standard expression used in DIS studies as a consistency check of normalizations.

With the momenta as in \cref{fig:momlabels}, we write the necessary dot products, using $q = k - k^\prime$ and $s + t + u = m_N^2 + m_\mQ^2 + \M^2$:
\begin{subequations}
        \label{eq:dot_prods}
   \begin{align}
        2\,\sp{p}{k} &= s- m_N^2~, \\
        2\,\sp{k}{q} &= t - m_\mQ^2~, \\
        2\,\sp{p}{q} &= \M^2 - m_N^2 - t ~. 
    \end{align}
\end{subequations}

The Lorentz-invariant differential cross section for the reaction 
$\gamma p \to \mQ \, \mX$, is given by~\cite{JPAC:2021rxu}: 
\begin{widetext}
\begin{equation}
    \label{eq:sumoverX}
 E_\mQ \frac{d^3\sigma}{d^3 k'} = \frac{1}{16\pi^3}\frac{1}{4E_\gamma \sqrt{s}} \, \frac{1}{4}\sum_{\mN} \sum_{\{\lambda\}}  \,  \int \prod_n \frac{d^3p_n }{(2\pi)^3\,2E_n} \, \left|A^{\gamma N \to \mQ\mX}_{\{\lambda\}}\right|^2 (2\pi)^4  \, \delta^4\!\left(k+p - k' - \sum_n p_n \right)~,
\end{equation}
\end{widetext}
where  the sum over $\mN$ runs over all possible final states containing $n$ unobserved particles, and \mbox{$\{\lambda\} = \lambda_\gamma, \lambda_N, \lambda_\mQ,\lambda_1,\lambda_2,\dots,\lambda_n$} collectively denotes the particle helicities. We recall that $E_\gamma = (s - m_N^2)/2\sqrt{s}$ is the photon beam energy in the center-of-mass frame. If the process is dominated by photon exchange, the amplitude factorizes into a top vertex, which describes the $\gamma \to \mQ \gamma^*$ interaction, and a bottom matrix element that encodes the $\gamma^* N \to \mX$ piece:
\begin{equation}
    A^{\gamma N \to \mQ\mX}_{\{\lambda\}} =   \mT_{\lambda_\gamma \lambda_\mQ}^\mu  \, g_{Q\gamma\gamma} \left(\frac{-g_{\mu\nu}}{t}\right)  e \, \mB_{\lambda_N \{\lambda_n\}}^\nu~,
\end{equation}
where $\mT_{\lambda_\gamma \lambda_\mQ}^\mu = \mel{\lambda_\gamma}{J^\mu(0)}{ \lambda_{\mQ} }$ is given in \cref{eq:mT}, and
\begin{equation}
 \mB_{\lambda_N \{\lambda_n\}}^\nu =  \mel{\lambda_N}{J_\nu(0)}{ \{\lambda_{n}\} }~.
\end{equation}
The same decomposition may be done in DIS, where the photon beam and quarkonium {\mQ} are replaced by an electron beam and a recoiling electron,
\begin{equation}
    A^{e N \to e'\mX}_{\{\lambda\}} = \ell_{\lambda_e \lambda_{e'}}^\mu \, e \left(\frac{-g_{\mu\nu}}{t}\right)  e \, \mB_{\lambda_N \{\lambda_n\}}^\nu~,
\end{equation}
where $\ell^\mu_{\lambda_e,\lambda_{e'}} = \mel{\lambda_e}{J^\mu(0)}{ \lambda_{e'} } = \bar{u}(k^\prime,\lambda_{e'}) \, \gamma^\mu \, u(k, \lambda_e)$.

One can thus rewrite \cref{eq:sumoverX} as
\begin{equation}
 E_\mQ \, \frac{d^3\sigma}{d^3 k'} = \frac{1}{4\pi^2}\frac{g_{\mQ \gamma\gamma}^2 \, e^2}{4E_\gamma \sqrt{s}} \,  \mT_{\mQ\gamma}^{\mu\nu} \,|\mP_\gamma|^2 \, W_{\mu\nu}~,
\end{equation}
where $\mT_{\gamma\mQ}^{\mu\nu} = \frac{1}{2}\sum_{\lambda_\gamma\lambda_\mQ} T^\mu_{\lambda_\gamma \lambda_\mQ}T^{*\nu}_{\lambda_\gamma \lambda_\mQ}$ as given in \cref{eq:mT_tensor}, and
\begin{widetext}
    \begin{align}
        W_{\mu\nu} &= \frac{1}{4\pi} \times
        \frac{1}{2} \sum_{\mN} \sum_{\lambda_N\{\lambda_n\}}  \,  \int \prod_n \frac{d^3p_n }{(2\pi)^3\,2E_n} \,  (2\pi)^4  \, \delta^4\!\left(k+p - k' - \sum_n p_n \right)\,
        \mel{\lambda_N}{J_\mu(0)}{ \{\lambda_{n}\} } \mel{\{\lambda_{n}\}}{J_\nu(0)}{ \lambda_N }\nonumber\\
&=\frac{1}{4\pi} \times \frac{1}{2} \sum_{\lambda_N}\, \int d^4z \, e^{i q\cdot z} \, \mel{\lambda_N}{\left[J_\mu(z), J_\nu(0)\right]}{\lambda_N}  ~,
    \end{align}
\end{widetext}
which in terms of structure functions yields \cref{eq:W_def}. 
We thus get
\begin{align}
    \frac{d^2\sigma}{dt \, dM_\mX^2} &= \frac{\pi}{(2\sqrt{s}\, E_\gamma)} \, E_\mQ \, \frac{d^3\sigma}{d^3k'} \nonumber\\&= \frac{1}{8\pi} \frac{g_{\mQ \gamma\gamma}^2 \, e^2}{(2\sqrt{s}E_\gamma)^2} \,  \mT_{\gamma\mQ}^{\mu\nu} \, |\mP_\gamma|^2 \, W_{\mu\nu}~,
\end{align}
which was given in \cref{eq:d2sigma}.

This normalization convention coincides with the one adopted for DIS in Ref.~\cite{ParticleDataGroup:2022pth}, where the elastic $e^-p$ double-differential cross section is given as:
    \begin{equation}
        \label{eq:d2sig_PDG}
         \frac{d^2\sigma}{d\xB \, dy} = \frac{y}{8\pi} \, L^{\mu\nu}  \left(\frac{e^2}{Q^2}\right)^2 W_{\mu\nu} ~,
    \end{equation}
in terms of $Q^2 = -t$ and the dimensionless variables $\xB$ given by \cref{eq:xb} and $y = \sp{p}{q}/\sp{p}{k}$, the fraction of energy lost by the electron beam in the lab frame (and not the fraction of transverse momentum defined in Ref.~\cite{Winney:2022tky}). 
The lepton tensor is defined as:
    \begin{equation}
        L^{\mu\nu} = \frac{1}{2} \sum_{\lambda,\lambda^\prime} \ell^{\mu}_{\lambda,\lambda^\prime} \, \ell^{\dagger \nu}_{\lambda,\lambda^\prime} ~.
    \end{equation}
Using \cref{eq:dot_prods}, we can see that this can be expressed in terms of $t$ and $\M^2$ as:
    \begin{equation}
        \label{eq:d2sig_jacobians}
         \frac{d^2\sigma}{dt \, d\M^2} = \frac{1}{(2\sqrt{s} \, E_\gamma)^2} \frac{1}{y} \frac{d^2\sigma}{d\xB \, dy} ~,
    \end{equation}
This mirrors the definition \cref{eq:mT_tensor} except with the coupling factored out in \cref{eq:d2sig_PDG} and thus replacing \mbox{$e^2 \, L^{\mu\nu} \to g_{Q\gamma\gamma}^2\,\mT^{\mu\nu}_{\gamma\mQ}$} in \cref{eq:d2sig_jacobians} yields \cref{eq:d2sigma}.


\bibliographystyle{apsrev4-1.bst} 
\bibliography{bibs}
\end{document}